\documentclass[]{aa}
\usepackage{graphicx}
\usepackage[varg]{txfonts}
\usepackage{natbib}
\usepackage[fleqn]{amsmath}
\usepackage[modulo, switch]{lineno}
\usepackage{nicefrac}
\usepackage{textcomp}
\usepackage{xfrac}
\usepackage[scientific-notation=true]{siunitx}
\usepackage{hyperref}
\usepackage{xcolor}
\usepackage{txfonts}
\def\d{\,d$^{-1}$}

\def\teff{$T_{\mathrm{eff}}$}
\def\lg{\ensuremath{\log g}}

\def\sun{\hbox{$_\odot$}}
\def\hd{HD\,201433}
\def\bc{BRITE\,-\,Constellation}
\def\smei{\textit{SMEI}}
\newcommand{\vs}{$v_{\rm e}\sin i$}

\newcommand{\Vmac}{$V_{\rm mac}$}

\newcommand{\Hermes} {\textsc{Hermes}\xspace}

\begin{document}
%\linenumbers
%
%Angular momentum transfer onto the surface of the SPB star HD\,201433 as seen by \bc\
%The SPB star \hd\ observed by \bc\
   \title{Triple system HD\,201433 with a SPB star component seen by \bc \thanks{Based on data collected by the \bc\  satellite mission, built, launched and operated thanks to support from the Austrian Aeronautics and Space Agency and the University of Vienna, the Canadian Space Agency (CSA), and the Foundation for Polish Science \& Technology (FNiTP MNiSW) and National Science Centre (NCN), the \Hermes spectrograph mounted on the 1.2\,m Mercator Telescope at the Spanish Observatorio del Roque de los Muchachos of the Instituto de Astrof{\'i}sica de Canarias, and the \textit{Solar Mass Ejection Imager}, which is a joint project of the University of California San Diego, Boston College, the University of Birmingham (UK), and the Air Force Research Laboratory.}: Pulsation, differential rotation, and angular momentum transfer}

   \author{T. Kallinger\inst{1}
	  \and
	W. W. Weiss\inst{1}
	\and
	P. G. Beck\inst{2,3,4}
	\and
	A. Pigulski\inst{5}
	\and
	R. Kuschnig\inst{1,6}
	\and
	A. Tkachenko\inst{7}
	\and
	Y. Pakhomov\inst{8}
	\and
	T. Ryabchikova\inst{8}
	\and
	T. L\"uftinger\inst{1}
	\and
	P.\,L.~Palle,\inst{3,4}
	\and
	E. Semenko\inst{9}
	\and
	G. Handler\inst{10}
	\and
	O. Koudelka\inst{6}
	\and
	J. M. Matthews\inst{11}
	\and
	A. F. J. Moffat\inst{12}
	\and
	H. Pablo\inst{12}
	\and
	A. Popowicz\inst{13}
	\and
	S. Rucinski\inst{14}
	\and
	G. A. Wade\inst{15}
	\and
	K. Zwintz\inst{16}
            }

   \offprints{thomas.kallinger@univie.ac.at}

   \institute{
Institute for Astrophysics, University of Vienna, T\"urkenschanzstrasse 17, 1180 Vienna, Austria
	\and
Laboratoire AIM, CEA/DRF CNRS - Universit\'e Denis Diderot IRFU/SAp, 91191 Gif-sur-Yvette Cedex, France
         \and
Instituto de Astrof\'{\i}sica de Canarias, E-38200 La Laguna, Tenerife, Spain
	\and
Departamento de Astrof\'{\i}sica, Universidad de La Laguna, E-38206 La Laguna, Tenerife, Spain
	\and
Instytut Astronomiczny, Uniwersytet Wroc\l{}awski, Kopernika 11, 51-622 Wroc\l{}aw, Poland
	\and
Institut f\"ur Kommunnikationsnetze und Satellitenkommunikation, Technical University Graz, Inffeldgasse 12, 8010 Graz, Austria
	\and
Instituut voor Sterrenkunde, K.U. Leuven, Celestijnenlaan 200D, 3001 Leuven, Belgium	
	\and
Institute of Astronomy, Russian Academy of Sciences, Pyatnitskaya 48, 119017 Moscow, Russia
	\and
Special Astrophysical Observatory, Russian Academy of Sciences, 369167, Nizhnii Arkhyz, Russia
	\and
Nicolaus Copernicus Astronomical Center, ul. Bartycka 18, 00-716 Warsaw, Poland
	\and
Department of Physics and Astronomy, University of British Columbia, Vancouver, BC V6T1Z1, Canada
	\and	
D\'epartement de physique and Centre de Recherche en Astrophysique du Qu\'ebec (CRAQ), Universit\'e de Montr\'eal, CP 6128, Succ. Centre-Ville, Montr\'eal, Qu\'ebec, H3C 3J7, Canada
	\and
Institute of Automatic Control, Silesian University of Technology, Akademicka 16, 44-100 Gliwice, Poland
	\and
Department of Astronomy \& Astrophysics, University of Toronto, 50 St. George Street, Toronto, Ontario, M5S 3H4, Canada
	\and
Department of Physics, Royal Military College of Canada, PO Box 17000, Station Forces, Kingston, Ontario, K7K 7B4, Canada
	\and
Institut f\"ur Astro- und Teilchenphysik, Universit\"at Innsbruck, Technikerstrasse 25/8, 6020 Innsbruck, Austria
}

   \date{Received 15 February 2017 / Accepted 28 March 2017}

\abstract
%Context
{Stellar rotation affects the transport of chemical elements and angular momentum and is therefore a key process during stellar evolution, which is still not fully understood. This is especially true for massive OB-type stars, which are important for the chemical enrichment of the universe. It is therefore important to constrain the physical parameters and internal angular momentum distribution of massive OB-type stars to calibrate stellar structure and evolution models. Stellar internal rotation can be probed through asteroseismic studies of rotationally split non radial oscillations but such results are still quite rare, especially for stars more massive than the Sun. The slowly pulsating B9V star \hd\ is known to be part of a single-lined spectroscopic triple system, with two low-mass companions orbiting with periods of about 3.3 and 154\,days. 
}
%Aims
{Our goal is to measure the internal rotation profile of \hd\ and investigate the tidal interaction with the close companion.}
%Methods
{We used probabilistic methods to analyse the \bc\ photometry and radial velocity measurements, to identify a representative stellar model, and to determine the internal rotation profile of the star.}
%Results
{Our results are based on photometric observations made by \bc\ and the \textit{Solar Mass Ejection Imager} on board the Coriolis satellite, high-resolution spectroscopy, and more than 96 years of radial velocity measurements. We identify a sequence of nine frequency doublets in the photometric time series, consistent with rotationally split dipole modes with a period spacing of about 5030\,s. We establish that \hd\ is in principle a solid-body rotator with a very slow rotation period of 297$\pm$76 days. Tidal interaction with the inner companion has, however, significantly accelerated the spin of the surface layers by a factor of approximately one hundred. The angular momentum transfer onto the surface of \hd\ is also reflected by the statistically significant decrease of the orbital period of about 0.9\,s during the last 96 years.
}
%Conclusions
{Combining the asteroseismic inferences with the spectroscopic measurements and the orbital analysis of the inner binary system, we conclude that tidal interactions between the central SPB star and its inner companion have almost circularised the orbit. They have, however, not yet aligned all spins of the system and have just begun to synchronise rotation.}

\authorrunning{Kallinger et al.}
\titlerunning{HD201433}
\maketitle

\section{Introduction}	\label{sec:intro}
Massive stars are important for the chemical enrichment of the universe. Slowly pulsating B (SPB) stars are not amongst the most massive stars, but they share a similar internal structure and are therefore ideal to improve our understanding of massive stars.

Slowly pulsating B stars (SPB) were introduced to the zoo of variable stars by \cite{Waelkens1991}. They are non-radial multi-periodic oscillators on the main sequence between spectral type B3 and B9, with an effective temperature ranging from about 11,000 to 22,000 K, and a mass between 2.5 and 8\,M\sun\ \citep[e.g.][]{aerts2010}. They oscillate in high-order gravity (g) modes with frequencies typically ranging from 0.5 to 2\d , which are driven by the $\kappa$-mechanism acting due to the iron-group element opacity bump \citep[e.g.][]{Dziembowski1993}. Consecutive radial order $n$ gravity modes of the same spherical degree $l$ are expected to be equally spaced in period, and deviations from this regular pattern carry information about physical processes in the near-core region \citep[e.g.][]{miglio2008}.

%________________________________________________________________Tab. Obs.
\begin{table*}[t]
\begin{small}
\begin{center}
\caption{Overview of the  photometric observations of \hd\ obtained with BRITE-Toronto$^2$, BRITE-Lem$^3$, and the Coriolis/\smei\ satellite. The last three columns give the number of data points of the raw, reduced, and subsequently binned data set.
\label{tab:obs}}
\begin{tabular}{c|ccccccccc}
\hline
\noalign{\smallskip}
Satellite & Orbital period & Duty cycle & Cadence & HJD start & HJD end & Range in CCDT & raw & reduced & binned\\
&[min]&[\%]&[min]&\multicolumn{2}{c}{-2\,450\,000} &[\degr C] & \multicolumn{3}{c}{data points}\\
\noalign{\smallskip}
\hline
\noalign{\smallskip}
BTr &  98.2 &$\sim$16 &0.338&7\,184.66 &7\,340.63&4 -- 24 & 105\,326 &102\,339&4\,225\\
BLb &  99.6 &$\sim$10 &0.338&7\,273.78 & 7\,286.93&29 -- 39 & 1\,741& 1\,321& 87\\
\noalign{\smallskip}
\smei & 101.6 & $\sim$68 & 101.6 & 2\,675.44 & 5\,561.41 & -- & 33\,412 & 27\,863 & --\\
\noalign{\smallskip}
\hline
\end{tabular}
\end{center}
\end{small}
\end{table*}
%__________________________________________________________________

SPB stars are expected to be dominated by a convective core and a radiative envelope, and therefore experience internal mixing processes, which have a significant influence on the lifetime of the star by enhancing the size of the convective region in which mixing of chemical elements occurs. Such a mixing might be induced by convective core overshooting but also by internal differential rotation \citep[e.g.][]{aerts2003,dupret2004}. Despite their importance for realistic stellar structure and evolution models of massive stars, the physical details describing these processes are hardly known. This is mainly because of the small number of detailed investigations of SPB stars, partly due to the few identified modes in these studies \citep[for a recent review see][]{aerts2015}.

The Canadian space telescope MOST \citep{walker2003,matthews2004} was very successful in providing the high quality data of SPB stars that are necessary to challenge theory \citep[e.g.][]{walker2005,aerts2006,gruber2012,Jerzykiewicz2013}. The breakthrough in observing SPB stars came, however, with the \textit{Kepler} mission. Only recently, \cite{papics2014,papics2015} reported on the detection of a rotationally affected series of g-modes in the two SPB stars KIC\,7760680 and KIC\,10526294 that show clear signatures of chemical mixing and rotation and which enabled the first actual seismic modelling of SPB stars. A limitation in this respect is that \textit{Kepler} can only observe fairly faint stars, for which additional observational constraints (e.g., from high resolution spectroscopy or interferometry) are difficult to obtain.

\hd\ (HR 8094, V389 Cyg) is one of the brightest stars (V$\simeq$5.61\,mag) suspected to be a SPB star (due to its position in the Hertzsprung-Russel diagram). The B9V star is known to be member of a single-line spectroscopic triple system \citep{barlow1989}. The most recent determinations of \teff\ = 12193$\pm$360\,K, \lg\ = 4.24$\pm$0.2, and $v\sin{i}$ = 15\,km/s were published by \cite{takeda2014}. Its Hipparcos parallax is $8.64\pm0.55$\,mas \citep{leeuwen2007}, from which we obtain an absolute visual magnitude of $M_V=0.29\pm0.14$\, mag. Interpolation in the tables of \cite{Lejeune2001} for [Fe/H] = 0.0 (see Sec.\,\ref{sec:atmo_par}) indicates a bolometric correction of $BC_V=-0.70\pm0.05$. With $M_{bol,\sun}=4.76$\,mag \citep{kopp2011} we then obtain $L/L\sun=115\pm15$. These parameters locate HD\,201433 close to the cool border of the SPB domain.

In this paper we report high-precision photometric observations of \hd\ with \bc \footnote{\href{url}{http://www.univie.ac.at/brite-constellation/}}, which is an array of five nanosatellites devoted to high-precision, long-term photometry of bright stars as is described by \cite{weiss2014}. Our photometric analysis is primarily based on 156 days of BRITE-Toronto (BTr) observations supplemented by about 13 days of BRITE-Lem (BLb) data (see Tab.\,\ref{tab:obs}). The data products and necessary post-processing are described in Sec.\,\ref{sec:obs} \& \ref{sec:postproc}. The  Bayesian frequency analysis (Sec.\,\ref{sec:freqAna}) reveals a sequence of nine significant close pairs of frequencies, consistent with rotationally split dipole modes, from which we extracted an average period spacing and rotational splittings (Sec.\,\ref{sec:periodspacing}). In Sec.\,\ref{Sec:SMEI} we demonstrate that our interpretation of the BRITE photometry is fully consistent with the signal found in the almost eight-year long \smei\ observations. We then construct a dense  stellar model grid and search for a representative model of \hd\ (Sec.\,\ref{Sec:Seis}), which we use in Sec.\,\ref{Sec:Rot} to infer the internal rotation profile. To complement the space photometry we obtain new high-resolution spectra which extend the time base to slightly more than 96 years with a total of 231 spectra usable for an orbital analysis. Based on an entirely Bayesian analysis of the radial velocity measurements we  improve the published orbital elements and  find evidence for a continuously decreasing  orbital period. Putting this in context of our asteroseismic and spectroscopic results we conclude that the main component of \hd\ is a SPB star of about three solar masses showing no significant rotational gradient throughout most of its interior. However, we find indications for tidal interaction with a close companion, causing an acceleration of the outermost envelope. We discuss our findings for \hd\ in a broader astrophysical context in Sec.\,\ref{sec:discussion} and summarise our analysis in Sec.\,\ref{sec:summary}.

%--------------------------------------------------------------------
\begin{figure}[t]
	\begin{center}
	\includegraphics[width=0.5\textwidth]{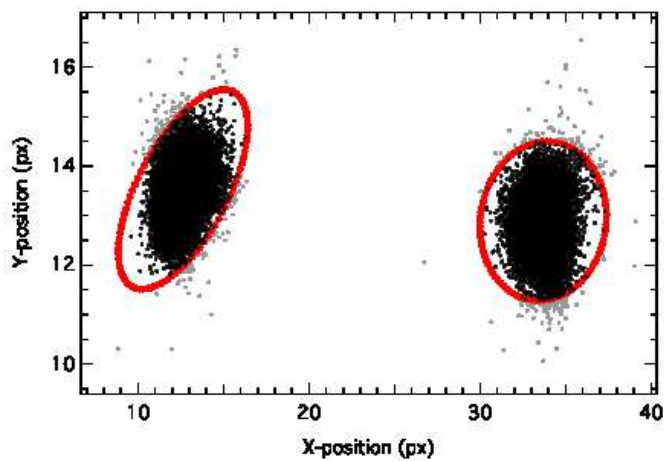}
	\caption{Point spread function (PSF) positions for the BTr observations (during observing setup 3) in the CCD subraster. The red ellipses indicate the limit outside of which data points are eliminated for further analysis (light grey points). We note that even though the ellipse on the left hand side appreas to  be misaligned it correctly reflects the distribution of the PSF positions.} 
	\label{fig:XYpos} 
	\end{center} 
\end{figure}
%--------------------------------------------------------------------

\section{BRITE photometry of \hd }	\label{sec:obs}

The photometric observations used in this study were carried out with two of the five \bc\ satellites. Each of the 20$\times$20$\times$20\,cm satellites hosts an optical telescope of 3\,cm aperture, feeding an uncooled CCD, and is equipped with a single filter. Three nanosats have a red filter (550--700\,nm) and two have a blue filter (390--460nm). The orbital periods are close to 100\,min, enabling continuous observations of the chosen target fields for about 5--30\,min per orbit. 

The detector is a Kodak KAI-11002M CCD with about 11 million 9 $\times$ 9\,$\mu$m pixels (plate scale of 27.3\arcsec\ per pixel), a 14-bit A/D converter, an inverse gain of about 3.5\,e$^-$/ADU, and a readout noise and dark current of about 16\,e$^-$ and 20\,e$^-$/s per pixel, respectively, at +20\degr\,C. %\citep{pablo2016}
The saturation limit of the pixels at this temperature is about 13\,000\,ADU, with the response being linear up to about 9\,000\,ADU. Further details about the detector and data acquisition of \bc\  are described by \cite{pablo2016}

A problem affecting the BRITE nanosatellites is the higher-than-expected sensitivity of the CCDs to particle radiation, which posed a major threat to the lifetime and effectiveness of the BRITE mission. The impact of high-energy protons causes the emergence of hot and warm pixels at a rate much higher than originally expected. The affected pixels more easily generate thermal electrons and thereby significantly impair the photometric precision of the observations. An additional important problem that appeared after several months of operation was the charge transfer inefficiency also caused by the protons. There were serious problems in the early phase of the mission, but thanks to slowing the readout time and adopting a \textit{chopping}  technique for data acquisition, the effect of CCD radiation damage on the photometry is now significantly reduced \citep{popowicz2016}. 

Satellite pointing is adjusted slightly between consecutive exposures in the chopping mode, so that the target PSFs alternate between two positions (about 20 pixels apart) on the CCD. This means that the PSF-free part of a given subraster image acts as a dark image for the subsequent exposure and subtracting consecutive exposures results in an image with one negative and one positive target PSF. The background defects are thereby almost entirely removed. More details about this technique are given by \cite{popowicz2016}.

\hd\ was one of the targets in the \bc\ Cygnus\,II field and was observed with BRITE-Toronto\footnote{The Canadian satellite BRITE-Toronto was launched on June 19, 2014, into a slightly elliptical and almost Sun-synchronous orbit and is equipped with a \textit{red} filter.} for about 156 days in June--November 2015 typically 48 times per BRITE orbit with an average cadence of 5\,s exposures every 20.3\,s. A significantly shorter data set was obtained with BRITE-Lem\footnote{The Polish BRITE-Lem was launched on September 21, 2013, into an elliptical orbit and is equipped with a \textit{blue} filter.}, which observed \hd\ for about 13 days in September 2015 for typically 30 times per orbit (see Tab.\,\ref{tab:obs}).

%--------------------------------------------------------------------
\begin{figure}[h]
	\begin{center}
	\includegraphics[width=0.5\textwidth]{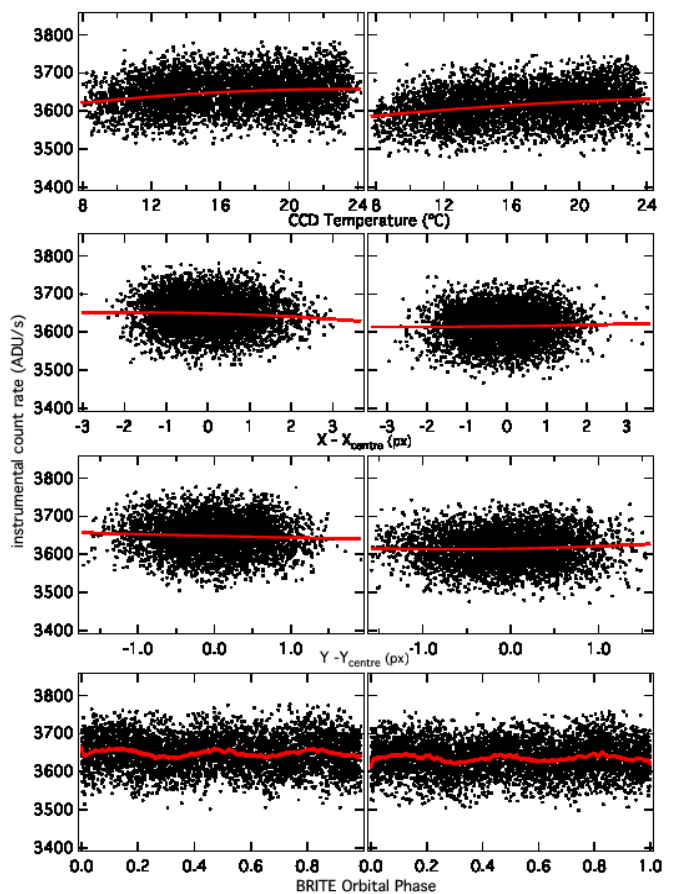}
	\caption{Correlations between \hd\ flux measurements and BTr housekeeping parameters (CCD temperature and the X and Y positions of the PSF) as obtained during observing setup 3. The left and right panels correspond to flux measurements extracted from the ``left'' and ``right'' part of the subraster image (see Fig.\,\ref{fig:XYpos}). The bottom panels show the residual time series phased with the satellite's orbital period. Red lines indicate linear (top panel) and polynomial (middle panels) fits and a boxcar filter (bottom panel).} 
	\label{fig:decor} 
	\end{center} 
\end{figure}
%--------------------------------------------------------------------

%--------------------------------------------------------------------
\begin{figure*}[t]
	\begin{center}
	\includegraphics[width=1.0\textwidth]{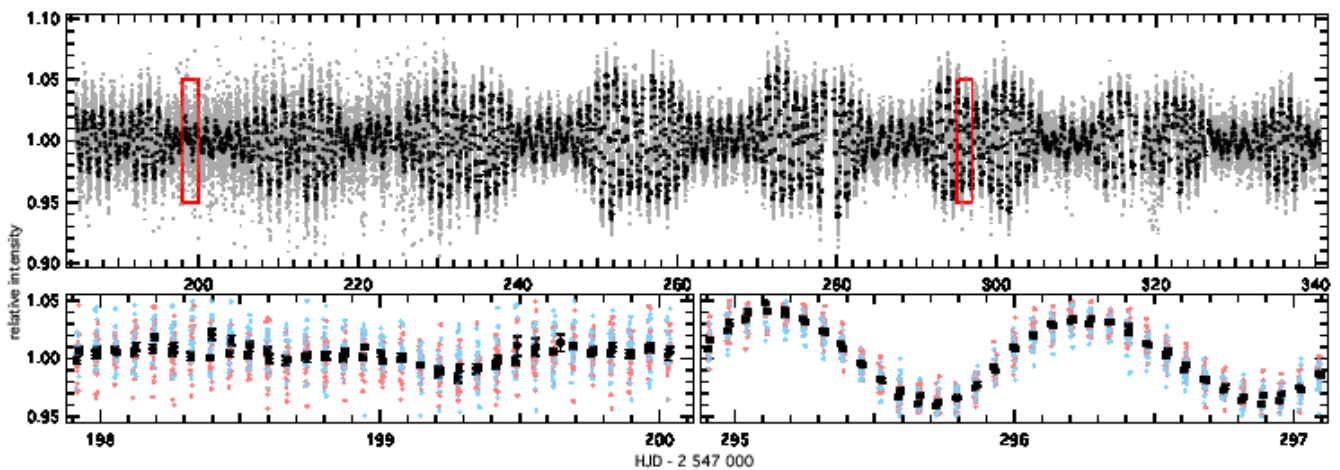}
	\caption{Final light curve of \hd\ as obtained with BTr. The grey and black dots in the top panel represent the full and binned data, respectively. The bottom panels show enlargements of the full data set (red boxes in the top panel) with the red and blue symbols indicating flux measurements extracted from the ``left'' and ``right'' part of the subraster frame (see Fig.\,\ref{fig:XYpos}). Black filled circles correspond to data binned into two bins per orbit.} 
	\label{fig:LCfinal} 
	\end{center} 
\end{figure*}
%--------------------------------------------------------------------

As the stellar flux is extracted from differential images (chopping mode) no bias, dark, and background corrections are necessary. The remaining main step is to identify the optimal apertures and to extract the flux within these apertures \citep{popowicz2016}. The light curves resulting from this pipeline reduction are deposited in the \bc\ data archive from where we extracted the data of \hd\ and applied some post-processing routines, as are described in the following.

\section{BRITE data post-processing}	\label{sec:postproc}

The raw stellar flux still includes instrumental effects and obvious outliers and therefore needs some post-processing. The BTr data came in two different setups\footnote{Setup refers here to a set of camera parameters but also to subraster positions on the CCD. Setups may change at the begining of a run (during optimising of the observations) or due to adding/removing a star during the run, where for each parameter change a whole new setup is generated with unique ID.} and five data blocks of approximately equal length, which we treated independently. The first setup at the beginning of the observing run addressed 24 stars in the field but had to be reduced to 18 stars (2nd setup), because of data transfer limitations. Subdivision of the dataset into blocks was required due to a limit of typically 30\,000 frames for the standard data reduction software.

Adapting the recipe of \cite{pigulski2016} we perform the following steps for each of the five blocks: 
\begin{itemize}
  \item Divide the data set into two sets corresponding to the alternating PSF position on the CCD subraster.  (see Fig.\,\ref{fig:XYpos}). 
  \item Compute a 2D histogram of the X/Y positions and fit a 3D multivariate Gaussian to it. Measurements that were obtained with the PSF centre positioned outside three times the widths of the Gaussian (see Fig.\,\ref{fig:XYpos}) are eliminated from further processing. This procedure identifies most of the outliers (about 2.5\% of the original data) and rejects them.
 \item The procedure resumes with the ``cleaned" data set and applies a 4$\sigma$-clipping to the whole data set, where $\sigma$ was determined from the complete set. The procedure results in the elimination of additional $\sim$0.3\%  of all data points.
\item To better access the instrumental correlations we first pre-whiten the two highest amplitude frequencies (see Sec.\,\ref{sec:freqAna}), which are subsequently added back after post-processing of the data.
 \item In the case of \hd\ the instrumental flux increases typically by 5 -- 7\,ADU/s per \degr C with increasing CCD temperature. A quadratic fit with the CCD temperature is sufficient to correct for this temperature correlation (see top panel of Fig.\,\ref{fig:decor}).
 \item Pixel-to-pixel sensitivity variations of the detector are reflected in correlations between the instrumental flux and the PSF position on the CCD (about 3 -- 5\,ADU/s per pixel). We correct for this with polynomial fits (middle panels of Fig.\,\ref{fig:decor}).
 \item Residual instrumental signal is apparent when phasing the instrumental flux with the satellite's orbital period. We correct for the high-overtone signal with a 200 point box-car filter in the phase plot (bottom panels of Fig.\,\ref{fig:decor}).
 \item The residual instrumental flux is then divided by its average value for conversion to relative flux.
\end{itemize}

The ten reduced data sets (two for each setup) are then simply stitched together, where no significant offsets at the subset interfaces are found. The final light curve of \hd\  consists of about 102\,000 individual measurements and is shown in Fig.\,\ref{fig:LCfinal}. The post-processing reduces the point-to-point scatter of the BTr data of \hd\ from about 46 to 34\,ADU/s (or $\sim$0.9\%).

The intrinsic variability of \hd\ acts on time scales of no shorter than a few hours; hence, the average cadence of about 20.3\,s (which corresponds to a Nyquist frequency of $\sim$2130\d ) is unnecessarily short. Given this and because the frequency analysis (see Sec.\,\ref{sec:freqAna}) requires good estimates for the uncertainties of the individual measurements we bin the light curve. To keep the dominant cadence short enough (i.e., the Nyquist frequency high enough) we bin the typically 48 measurements per BRITE orbit into two bins, where the standard deviation of the original measurements within a given bin provides a good estimate for the photometric accuracy. The binned light curve consists of about 4\,200 data points with a median cadence of $\sim$8.13\,min ($f_\mathrm{nyq} \simeq 89$\d ) and an average error of about 2.1\,ppt. % (see Tab.\,\ref{tab:obs}).

%--------------------------------------------------------------------
\begin{figure}[h]
	\begin{center}
	\includegraphics[width=0.5\textwidth]{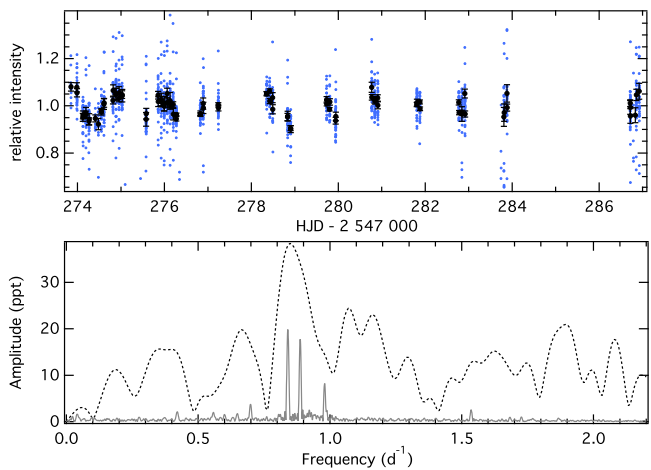}
	\caption{Final light curve (top) of \hd\ as obtained by BLb, with black and blue symbols corresponding to the binned and unbinned version, respectively. The bottom panel show the Fourier amplitude spectra of the binned BLb (black dashed line) and BTr (grey line) light curves.
} 
	\label{fig:BLb} 
	\end{center} 
\end{figure}
%--------------------------------------------------------------------

The BLb raw dataset of \hd\ has a considerably shorter time base than the BTr observations and is due to the higher CCD temperature also much noisier. The original $\sim$1\,700 measurements reduce to about 1\,300 useful data points in the post-processed light curve (see Fig.\,\ref{fig:BLb}). Binning of the typically 30 measurements per BRITE orbit results in 87 data points with a median cadence of $\sim$5.4\,min and an average error of about 16\,ppt.

%--------------------------------------------------------------------
\begin{figure}[h]
	\begin{center}
	\includegraphics[width=0.5\textwidth]{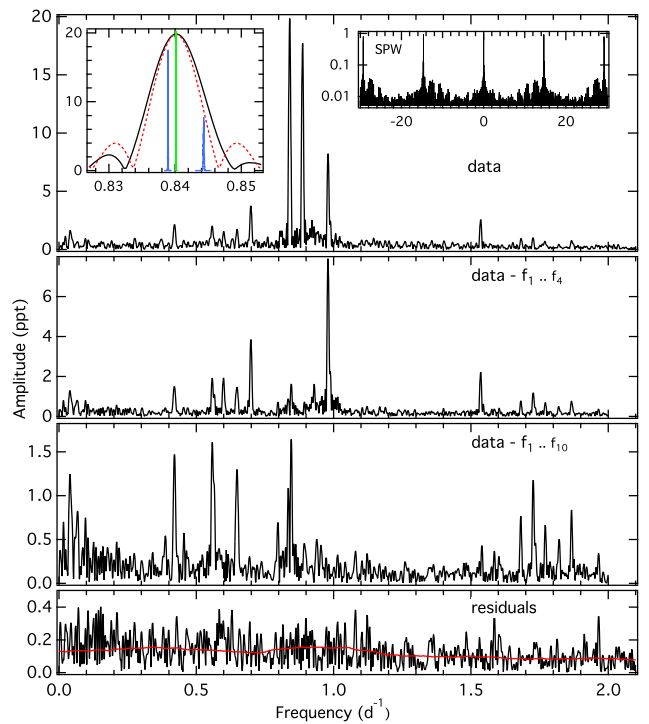}
	\caption{Fourier amplitude spectrum of the binned BTr light curve of \hd . The two middle panels show the amplitude spectrum after pre-withening of the given frequencies. The bottom panel gives the residual amplitude spectrum after pre-withening with all significant frequenices. The right insert in the top panel shows the spectral window function of the BTr dataset. The left insert gives the original spectrum (black line) and the spectral window (red dashed line) centered on the main peak. The green and blue peaks indicate the posterior parameter distributions (arbitralily scaled in amplitude for better visibility) of a single and multiple sine fit with \textsc{MultiNest}, respectively.}
	\label{fig:fspec} 
	\end{center} 
\end{figure}
%--------------------------------------------------------------------

\section{Frequency analysis} \label{sec:freqAna}

SPB stars are expected to show long-period g modes in a frequency range of up to a few cycles per day. This is well separated from residual instrumental signal and alias peaks due to the orbital frequency of BTr of $\sim$14.7\d\ (and multiples of it). We compute the Fourier amplitude spectrum of the unbinned light curve and find no significant peak between 2\d and the Nyquist frequency which cannot be attributed to the satellite's orbital frequency. The pulsation spectrum (see Fig.\,\ref{fig:fspec}) of \hd\ shows the strongest peaks between about 0.5 -- 1\d, and another group of peaks between about 1.5 -- 2\d. Above that no significant power can be found.

An unusual feature in the Fourier spectrum of \hd\ is that some peaks appear slightly broader than expected from the spectral window function (see insert in Fig.\,\ref{fig:fspec}). This indicates the presence of close frequencies that are separated by less than (or close to) the formal frequency (Rayleigh-) resolution of $1/T \simeq 0.0064$\d . Such features are problematic for a standard frequency analysis (based on a strict pre-whitening procedure) because assuming a mono-periodic signal in the vicinity of the considered Fourier peak yields a frequency that corresponds to the weighted average of the intrinsic frequency multiplet and pre-whitening this ``wrong'' signal causes artificial peaks in the spectrum. Furthermore, it is difficult (or often impossible) to objectively rate the significance of the result and its uncertainties. 

We use a probabilistic approach to tackle this problem. \cite{kal2016} have developed a fully automated Bayesian algorithm that searches for close frequencies in time series data and tests their statistical significance by comparison to a fit with constant (i.e., no periodic) signal and a fit with a mono-periodic signal. The procedure performs the following steps:
\begin{itemize}
	\item Compute the Fourier amplitude spectrum up to 2\d\ and determine the frequency with the highest amplitude.
	\item Fit $N$ functions, $F_{(t,n)} = \sum_{i=1}^n A_i \sin{[2\pi(f_it+\Phi_i)]} + c$, to the time series, where $n$ incrementally increases from 1 to $N$ so that, in total, $N$ models with 1, 2, ..., $N$ sinusoidal components are fit to the data. $A$, $f$, and $\Phi$ are the amplitude, frequency, and phase of the $i$th component, respectively. The parameter $c$ serves as an offset to ensure that $\int_{T}{F_{(t)}dt=0}$ even if the duration $T$ of the time series is not an integer multiple of the signal period. For the fit we use a Bayesian nested sampling algorithm \citep[\textsc{MultiNest;}][]{feroz2009}, and allow the individual frequencies to vary around the initial frequency by $\pm 2/T$, and the amplitudes between 0 and 50 times the initial amplitude from the amplitude spectrum. Phases have no initial constraints and can vary from 0 to 1. 
	\item To rate if a signal is statistically significant (i.e., not due to noise) and if so, which model best represents the data, we compute the model probability ($p_n$) by comparing the global evidences\footnote{The global evidence is a normalised logarithmic probability delivered by \textsc{MultiNest} describing how good the model fits the data with respect to the uncertainties, parameter ranges, and the complexity of the model.} ($z_n$) of the fits to those of a fit with a constant factor ($z_c$). If $p = \sum z_n/(z_c + \sum z_n) > 0.95$ we consider the solution as real\footnote{In probability theory an odds ratio of 10:1 (i.e., $p$=0.9) is considered already as strong evidence \citep{jeffreys98}.} and not to be due to noise. If so, the best-fit model is then the model with $p_n = z_n / \sum z_n > 0.95$. This means that in order to be accepted, a multiperiodic solution needs to fit the data considerably better than the monochromatic solution. Our approach for the statistical significance of a signal compares well to classical approaches like a SNR $>$ 4 \citep[e.g.][]{breger1993, kuschnig1997} but has the advantage of providing an actual statistical statement that is based only on the data and that allows us to discriminate between mono- and multi-periodic solutions for closely separated frequencies. In the present case we tested models with up to three components but find that for none of the identified multiplets is a solution with $N=3$ statistically significant.
	\item The best-fit parameters and their 1$\sigma$ uncertainties are then computed from the marginalised posterior distribution functions as delivered by \textsc{MultiNest}.
	\item The best-fit model is subtracted from the time series and the procedure starts from the beginning.
\end{itemize}
We stop the procedure when $p$ drops below 0.66 (corresponding to weak evidence) but we accept only those frequencies with $p>0.95$. We note that the frequency, amplitude, and phase uncertainties that are computed from the posterior probability distributions compare well with uncertainties determined from other criteria \citep[e.g.][]{kal2008}. 

Based on extensive tests with synthetic data (with the sampling and noise characteristics of the BTr data of \hd) \cite{kal2016} have shown that the algorithm is capable to reliably (>99.9\%) distinguish between a single frequency and a pair of close frequencies if the frequencies are separated by more than $\sim 0.5/T$ and their amplitudes are larger than about 1\,ppt. The uncertainties of the individual frequencies are thereby only slightly larger than for an unperturbed mono-periodic signal but rarely exceed $0.1/T$.

\subsection{Frequencies and frequency combinations in the BTr data}

Our Bayesian frequency analysis algorithm identified 9 ``features'' in the binned BTr data of \hd\ that consist of statistically significant closely separated frequencies in addition to a further 11 single frequencies. An example for a pair of close frequencies is illustrated in the left insert in Fig.\,\ref{fig:fspec}, where we show the posterior parameter distributions of the one-frequency and two-frequency model fits for the highest-amplitude peak in the BTr spectrum of \hd. The evidence of the two-frequency model is orders of magnitude better than for the one-frequency model (despite the Bayesian ``penalty'' for introducing additional free model parameters), which indicates -- based on solid statistical grounds -- that  more than one frequency is needed to reproduce the data in this frequency range. 

\begin{figure*}[t]
	\begin{center}
	\includegraphics[width=1.0\textwidth]{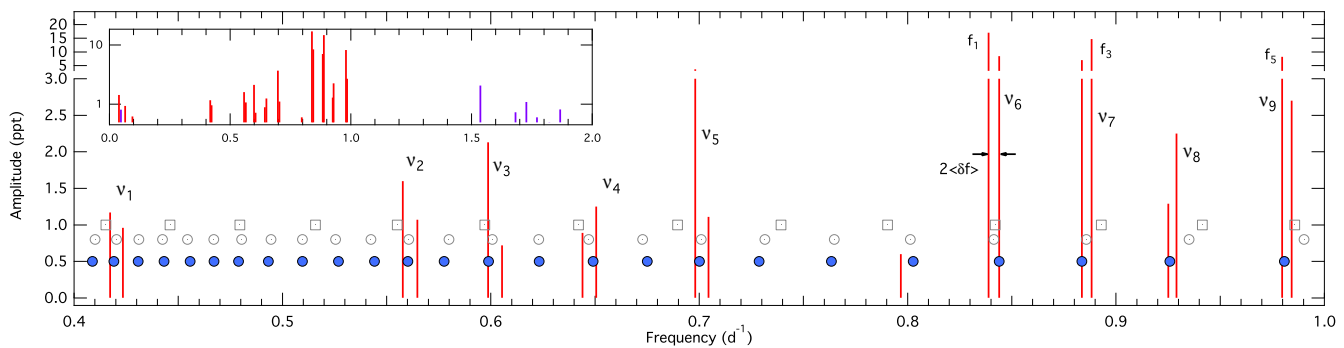}
	\caption{Schematic view of all significant frequencies in BTr observations of \hd . Grey squares and circles give two period spacings (see text and Fig.\,\ref{fig:Spacings}). The identified rotational doublets are labeled with $\nu_i$. Blue circles indicate the non-adiabatic frequencies of a representative MESA model. The insert shows the full frequency range, with independent frequencies in red and combination frequencies in blue.} 
	\label{fig:freq} 
	\end{center} 
\end{figure*}

The 29 significant frequencies detected in the BTr data set are listed in Tab.\,\ref{tab:freq}. After pre-whitening them from the data, the residual spectrum (see Fig.\,\ref{fig:fspec}) has an average amplitude of about 110\,ppm. The ``bump'' around 1\d, however, indicates that there is still some undetected signal left (which can well be of instrumental origin). We searched for linear combinations among all significant frequencies. Out of the 29 detected frequencies we find 22 independent frequencies. The remaining peaks correspond to first-order linear combination frequencies (where $f_i = f_j \pm f_k$). In order to be identified as a linear combination a frequency has to fulfil the criterion $(f_i - f_j \mp f_k)^2 < \sigma_{f_i}^2 + \sigma_{f_j}^2 + \sigma_{f_k}^2$ and its amplitude must be smaller than the amplitudes of its parent frequencies ($f_j$ and $f_k$). We also searched for higher-order combinations but found none. A schematic view of the independent and combination frequencies is shown in Fig.\,\ref{fig:freq} indicating that all peaks above 1\d\ are actually combination frequencies and that 18 of the 19 peaks between 0.4 and 1\d\ are part of a pair of close frequencies fully consistent with rotationally split dipole modes.

%--------------------------------------------------------------------
%--------------------------------------------------------------------

%________________________________________________________________Tab. Freq.
\begin{table}[t]
\begin{tiny}
\begin{center}
\caption{Significant frequencies in the BTr observations of \hd . Uncertainties for the frequency $f$, amplitude $A$, and phase $\Phi$ are given in parentheses in units of the last digit. The phase is defined for the beginning of the data set (mHJD  = 184.6694) The frequencies listed in the bottom part are combination frequencies, where $\epsilon$ gives the deviation between the observed frequency and the combination of its parental frequencies in units of the uncertainties (e.g., $\epsilon<1$ means a difference within 1$\sigma$ of the formal uncertainties). }
\label{tab:freq}
\begin{tabular}{lcccccc}
\hline
\noalign{\smallskip}
ID&	$f$ (d$^{-1}$) &	$A$ (ppt) &	$\Phi$ (1) &	$p$ &	SNR & $\epsilon$\\
\noalign{\smallskip}
\hline
\noalign{\smallskip}
$f_{1}$ &        0.83881(9)   &    16.99(38)       &    0.555(6)     &    1.000    &   34.4   &\\   
$f_{2}$ &        0.84398(17)  &      8.47(38)      &    0.192(13)   &    1.000   &   17.2   &\\   
$f_{3}$ &        0.88363(29)  &      7.04(64)      &    0.462(23)   &    1.000   &    19.1  &\\   
$f_{4}$ &        0.88831(14)  &    14.68(66)      &    0.077(12)   &    1.000   &    39.8  &\\   
$f_{5}$ &        0.97975(10)  &      8.22(17)      &    0.213(9)      &    1.000   &    31.6  &\\   
$f_{6}$ &        0.98431(29)  &      2.70(17)      &    0.576(26)   &     1.000  &    10.3  &\\   
$f_{7}$ &        0.69813(32) &       3.68(25)      &    0.885(25)   &     0.999  &    17.9  &\\   
$f_{8}$ &        0.7045(11)  &     1.11(24)       &    0.406(81)   &     0.999  &     5.4  &\\     
$f_{9}$ &        0.59867(30)  &     2.13(13)       &    0.539(26)   &     0.972  &    11.6  &\\   
$f_{10}$ &     0.60535(92)  &     0.724(12)     &    0.142(70)  &      0.972  &     3.9   &\\    
$f_{11}$ &     0.92511(63)  &     1.29(26)       &    0.833(53)   &     1.000  &      7.2  &\\    
$f_{12}$ &     0.92901(31)  &     2.25(25)       &    0.106(26)   &     1.000  &     12.6 &\\    
$f_{13}$ &     0.55775(38)  &     1.60(13)       &    0.686(35)   &     1.000  &      9.7  &\\    
$f_{14}$ &     0.56476(52)  &     1.07(13)       &    0.367(43)   &     1.000  &      6.5  &\\    
$f_{15}$ &     0.4173(11)   &     1.17(27)       &    0.87(10)      &    0.999  &      7.4  &\\    
$f_{16}$ &     0.4234(14)   &     0.96(26)       &    0.407(89)    &    0.999 &      6.1  &\\     
$f_{17}$ &     0.6440(13)  &     0.89(17)       &    0.435(95)   &     0.999 &      5.8  &\\     
$f_{18}$ &     0.65052(73)  &     1.25(18)       &   0.040(46)    &     0.999  &     8.2   &\\    
$f_{19}$ &     0.03933(46)  &     1.44(13)       &   0.873(43)    &     1.000  &     9.8   &\\    
$f_{20}$ &     0.06500(41)  &      0.94(13)      &   0.591(47)    &     1.000  &     6.4   &\\    
$f_{21}$ &     0.7967(12)      &      0.60(14)      &    0.469(11)    &     0.999  &    4.6    &\\
$f_{22}$ &     0.09564(83) &      0.62(13)      &    0.871(86)    &     0.999  &    4.6    &\\
\noalign{\smallskip}
\hline
\noalign{\smallskip}
$f_{1+7}$ &        1.53632(50)  &     2.07(13) &     0.421(22) &     1.000  &     10.8  &      1.03 \\  
$f_{1+4}$ &        1.72670(51)  &     1.09(13)  &     0.704(44) &     0.997  &       7.5  &      0.78 \\ 
$f_{3+6}$ &        1.86678(85)  &      0.82(13) &     0.479(51) &     1.000  &      6.0   &      1.23 \\ 
$f_{1+2}$ &        1.68194(71)  &      0.73(13) &     0.944(66) &     0.999  &       5.5  &      1.14 \\
$f_{1+6}$ &        1.8219(11)    &      0.49(12) &      0.512(93) &    0.999  &       4.1  &       1.07\\
$f_{7-18}$ &      0.04765(63) &      0.81(13) &      0.273(48) &    0.985  &       5.8  &       0.04 \\
$f_{3+4}$ &        1.77067(91) &      0.60(13) &      0.736(78) &     0.999 &       4.7  &       1.32 \\
\noalign{\smallskip}
\hline

\end{tabular}
\end{center}
\end{tiny}
\end{table}
%__________________________________________________________________

\subsection{Frequencies in the BLb data}

The BLb data set has a much shorter time base and is noisier than the BTr dataset. Consequently, the frequency analysis is more challenging  (see the amplitude spectrum in Fig.\,\ref{fig:BLb}). An independent analysis gives only one significant peak with a frequency of $0.852 \pm 0.003$\d, which represents a weighted average of $f_1$ and $f_4$ of the formally unresolved frequencies in BLb. We can, however, fix the frequency to the values determined for the BTr data and fit only the amplitude and phase. We tried various combinations of the three largest amplitude frequencies in the BTr data ($f_1$, $f_4$, $f_5$, $f_1\land f_4$, $f_1\land f_5$, $f_4 \land f_5$, and $f_1 \land f_4 \land f_5$) and find that a fit with $f_1$ and $f_4$ gives the (by far) the best model evidence. The resulting amplitudes and phases are given in Tab.\,\ref{tab:freq1}. The two frequencies have very similar amplitude ratios and phase differences in the two BRITE passbands, indicating that they have the same spherical degree \citep[e.g.][]{2008CoAst.152..140D}.

%________________________________________________________________Tab. Freq BLb.
\begin{table}[h]
\begin{tiny}
\begin{center}
\caption{Amplitude and phases of $f_1$ and $f_4$ in the BTr and BLb passbands and the corresponding amplitude ratios and phase differences.
\label{tab:freq1}}
\begin{tabular}{l|cc|cc|cc}
\hline
\noalign{\smallskip}
ID&\multicolumn{2}{c|}{BTr} &\multicolumn{2}{c|}{BLb} &&\\
&	$A$ (ppt) &	$\Phi$ (1) &	$A$ (ppt)  & $\Phi$ (1) & $A_b/A_r$ & $\Phi_b - \Phi_r$\\
\noalign{\smallskip}
\hline
\noalign{\smallskip}
f$_{1}$          &   16.99(38)   &    0.555(6)  &    30.2(3)    &    0.579(14)   &   1.78(4)   &   0.02(2) 	\\ 
f$_{4}$          &   14.68(66)   &    0.08(1)  &    24.6(3)    &    0.198(16)   &   1.69(8)   &   0.12(2) 	\\ 
\noalign{\smallskip}
\hline
\end{tabular}
\end{center}
\end{tiny}
\end{table}
%__________________________________________________________________

\section{Period spacings and rotational splittings}	\label{sec:periodspacing}

\begin{figure}[t]
	\begin{center}
	\includegraphics[width=0.5\textwidth]{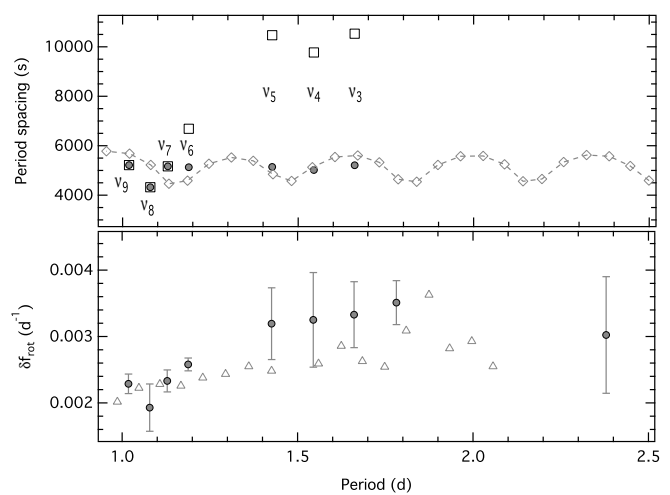}
	\caption{Measured period spacings (top) and rotational split frequencies (bottom). The period spacings are shown for two different solutions (see text). While for $S_1$ (open squares) $dP_i = 1/\nu_{i-1} - 1/\nu_{i}$ is plotted as observed (only for $\nu_6$ the measured spacing is divided by 3), for $S_2$ (filled circles) we divide the measured spacing for $\nu_{3-5}$ by two and for $\nu_6$ by 4 ($\nu_{7-9}$ are the same as for $S_1$). Diamond symbols (connected with a dashed line) represent period spacings of a representative MESA model. Triangle symbols in the bottom panel give the rotational splittings for KIC\,10526294 \citep{papics2014}, where we exclude the two measurements with the largest uncertainties (for better visibility).} 
	\label{fig:Spacings} 
	\end{center} 
\end{figure}

From our list of significant independent frequencies we can identify 9 rotationally split doublets ($\nu_1$ -- $\nu_9$ in Fig.\,\ref{fig:freq}), which indicates that all modes have the same spherical degree of $l=1$. Our mode identification is summarised in Tab.\,\ref{tab:split}, where we assume symmetric triplets with the unobserved (i.e., unresolved) central ($m = 0$) component being located half way between the $|m| = 1$ components. There are obviously some modes missing so that measuring the period spacing is not straightforward. If we define the period spacing as $dP_i = 1/\nu_{i-1} - 1/\nu_{i}$ and assume that between $\nu_5$ and $\nu_6$ two modes are missing (i.e., $dP_6 \simeq [1/\nu_5 - 1/\nu_6]/3$) we find the period spacings presented in Fig.\,\ref{fig:Spacings} as solution $S_1$. There is a strong gradient of $dP$ with the period, which is not consistent with theory \citep[e.g.][]{miglio2008}. A more realistic solution ($S_2$ in Fig.\,\ref{fig:Spacings}) is found when assuming 3 modes missing between $\nu_5$ and $\nu_6$ and one mode missing between $\nu_2$ -- $\nu_3$, $\nu_3$ -- $\nu_4$, and $\nu_4$ -- $\nu_5$. The resulting period spacings are quite uniform with only small deviations from the median value of about 5030\,s, which is a typical value for a star like \hd\ \citep[see KIC\,10526294;][]{papics2014}. Our assumption of filling missing modes might appear unrealistic but in fact alternating high- and low-amplitude modes were already observed in KIC\,10526294, so that it is not surprising that some modes between detected modes fall below the detection limit of our observations ($\sim$0.5\,ppt). In Fig.\,\ref{fig:freq} we show predictions of the mode sequences based on both solution but note that we cannot distinguish between them based on the observations alone.

%--------------------------------------------------------------------
%--------------------------------------------------------------------

The average rotational splittings are determined as <$\delta \mathrm{f}_{rot}$>$ = (\nu_{+1} - \nu_{-1})/2$ (with the indices of $\nu$ indicating the $m$ value of the mode). This means that we again assume symmetric triplets, which contradicts the findings of \cite{papics2014}, but given the missing $m=0$ components this is the only possibility we have. Our splittings are, however, similar to what was reported for KIC\,10526294 and have an average value of $\Delta f=0.0028$\d . Even the trend of increasing splittings towards higher periods can be observed (bottom panel in Fig.\,\ref{fig:Spacings}), which implies a non-rigid internal rotation profile of the star. For solid body rotation the rotational splitting of $l=1$ gravity modes is equal to half the rotation rate of the star, so that we can estimate the average rotation period of \hd\ to be about 177\,d. Since the assumption of a solidly rotating star is likely wrong, this value represents the average rotation period dominated by the near-core region, where the g modes have the largest contribution to the splittings.

%________________________________________________________________Tab. RotSplittings.
\begin{table}[t]
\begin{small}
\begin{center}
\caption{Rotationally split triplets based on BTr and \smei\ observations, In the case of BTr, the unresolved central ($m=0$) component f$_0$ is determined from the midpoint of the observed f$_{\pm 1}$ components and P$_0$ is the corresponding period. The split frequency <$\delta \mathrm{f}_{rot}$> is then half the difference between the $m=1$ and $m=-1$ component and $n_g$ gives the radial order of the best-fit model frequency (see Sec.\,\ref{Sec:Seis}). In the case of \smei\ (labeled by an arrow in the first column), f$_0$ and <$\delta \mathrm{f}_{rot}$> are determined from a Lorentzian fit to the data. Frequencies and periods are given in units of \d\ and d, respectively.
\label{tab:split}}
\begin{tabular}{ccrrrrr}
\hline
\noalign{\smallskip}
ID &ID$_{obs}$& $m$ & $\mathrm{f}_{m}$ & $\mathrm{P}_{0}$ & $<$$\delta \mathrm{f}_{rot}$$>$& $n_g$\\
%&\multicolumn{2}{c|}{BTr} &\multicolumn{2}{c|}{BLb} &&\\
\noalign{\smallskip}
\hline
\noalign{\smallskip}
&$f_{15}$ 	& -1  & 0.4173(11) &  & &    \\
$\nu_{1}$ &  & 0 & 0.42031(89) & 2.3997(51)&0.00302(88)&    39\\
&$f_{16}$ 	&  1 & 0.4234(14) &&&\\
\noalign{\smallskip}
\hline
\noalign{\smallskip}
&$f_{13}$ 	& -1 & 0.55775(38) & & &     \\
$\nu_{2}$ &&0&0.56125(32) & 1.7818(10)&0.00351(33)&     29\\
&$f_{14}$ 	&  1 & 0.56476(52) &&&\\
\noalign{\smallskip}
\hline
\noalign{\smallskip}
&$f_{9}$ 	& -1 & 0.59867(30) & & &    \\
$\nu_{3}$&&0&0.60201(50) & 1.6611(14)&0.00333(50)&    27\\
&$f_{10}$ 	&  1 & 0.60535(92) &&&\\
\noalign{\smallskip}
\hline
\noalign{\smallskip}
&$f_{17}$ 	& -1 & 0.6440(13) & & &  \\
$\nu_{4}$&&0&0.64727(75) & 1.5449(18)&0.00325(71)&  25\\
&$f_{18}$ 	&  1 & 0.65052(73) &&&\\
\noalign{\smallskip}
\hline
\noalign{\smallskip}
&$f_{7}$ 	& -1 & 0.69813(32) & & &     \\
$\nu_{5}$&&0&0.70132(56) & 1.4259(11)&0.00319(54)&     23\\
&$f_{8}$ 	&  1 & 0.7045(11) &&&\\
\noalign{\smallskip}
\hline
\noalign{\smallskip}
&$f_{1}$ 	& -1 & 0.83881(9) & & &\\
$\nu_{6}$&&0& 0.84139(10) & 1.1885(1)&0.00258(10)&19\\
&$f_{3}$ 	&  1 & 0.84398(17) &&&\\
\noalign{\smallskip}
$\rightarrow$&		&0	&0.84086(16) & 1.1893(2)&0.00246(15)&\\
\noalign{\smallskip}
\hline
\noalign{\smallskip}
&$f_{3}$ 	& -1 & 0.88363(29) && &\\
$\nu_{7}$&&0&0.88597(16) &1.1287(2) &0.00233(16)&18\\
&$f_{4}$ 	&  1 & 0.88831(14) &&&\\
\noalign{\smallskip}
$\rightarrow$&		&0	&0.88578(13) &1.1289(2)& 0.00243(12)&\\
\noalign{\smallskip}
\hline
\noalign{\smallskip}
&$f_{11}$ 	& -1 & 0.92511(63) & & &\\
$\nu_{8}$&&0&0.92706(36) & 1.0787(4)&0.00193(36)&17\\
&$f_{12}$ 	&  1 & 0.92901(31) &&&\\
\noalign{\smallskip}
\hline
\noalign{\smallskip}
&$f_{5}$ 	& -1 & 0.97975(10) & & &\\
$\nu_{9}$&&0&0.98203(15) & 1.0183(2)&0.00229(15)&16\\
&$f_{6}$ 	&  1 & 0.98431(29) &&&\\
\noalign{\smallskip}
$\rightarrow$&		&0	&0.98164(15) &1.0187(2)& 0.00257(16)&\\
\noalign{\smallskip}
\hline
\end{tabular}
\end{center}
\end{small}
\end{table}
%__________________________________________________________________

%--------------------------------------------------------------------
\begin{figure}[t]
	\begin{center}
	\includegraphics[width=0.5\textwidth]{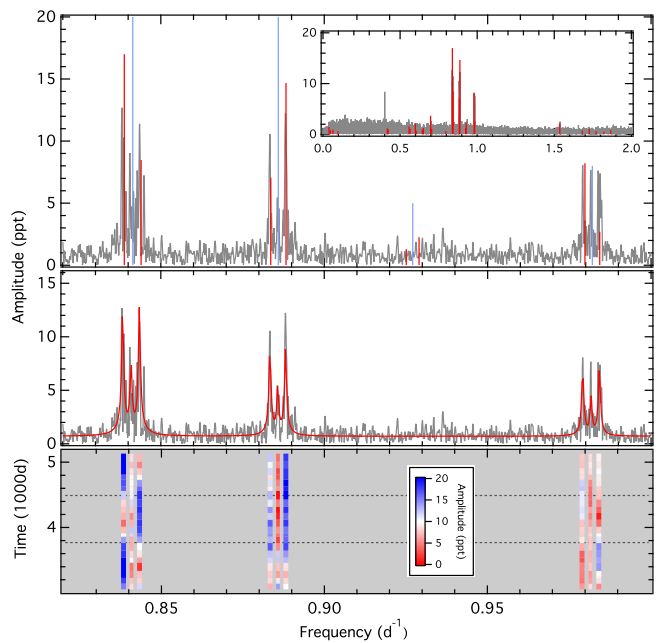}
	\caption{Fourier amplitude spectrum of the \smei\ data set of \hd . The \textit{top} panel gives the original spectrum (grey line) with the red and blue lines indicating the significant frequenices in the BTr data set (Tab.\,\ref{tab:freq}) and their mid-point, respectively. The insert shows the full frequency range. The \textit{middle} panel shows the same spectrum along with a sequence of Lorentzian profiles (red line) fitted to the spectrum. The \textit{bottom} panel shows the (color-coded) amplitude at the central frequencies of the Lorentzians as a function of time. The amplitudes are determined within a 720\,d long subset moved across the full time series in 100\,d steps. The horizontal dotted lines separate fully independent parts.}
	\label{fig:smei} 
	\end{center} 
\end{figure}
%--------------------------------------------------------------------

\section{Coriolis/\smei\ photometry of \hd} \label{Sec:SMEI}

Even though the frequency doublets identified in the BTr data set represent a statistically solid result, as is demonstrated in Sec.\,\ref{sec:freqAna}, the separations of the individual components are close to the formal frequency resolution of the time series. Such a result might be considered at first glance questionable as it disagrees with the \cite{Loumos1978} criterion, which requires two close peaks to have a minimal separation of $\sim$1.5 times the frequency resolution to avoid influence on their apparent frequencies when applying the classical Fourier technique. In our case we have the fortunate situation that we can test our claims with a data set long enough to provide the required frequency resolution. These data are provided by the \textit{Solar Mass Ejection Imager} (\smei ) on board the Coriolis satellite \citep{Eyles2003, Jackson2004}. 

\smei\ comprises three wide-field cameras, which are aligned such that the total field of view is a $180\deg$ and about $3\deg$ wide arc, so that a near-complete image of the sky is obtained after  about every 102\,min orbit. A detailed description of the data analysis pipeline used to extract light curves from these data is provided by \cite{Hick2007}. The data of \hd\ were taken from the \smei\ website\footnote{\href{url}{http://smei.ucsd.edu/new\_smei/index.html}} and contain about 33\,400 measurements covering almost eight years of near-continuous observations. The classical frequency resolution therefore is ten times better than the frequency splittings which we are discussing for \hd\ (see Tab.\,\ref{tab:split}). Details about the data set are given in Tab.\,\ref{tab:obs}. 

The \smei\ photometric time series are subject to strong instrumental effects, such as large yearly flux fluctuations, which obviously are due to an insufficient background correction. As a remedy we phased the data with a period of one year and computed the median flux in 200 phase sub-intervals and find an annual amplitude of about 0.8\,mag. We applied an Akima spline fit and computed the residuals to the smooth one-year variation. Back in the time domain one sees outliers, jumps, irregular intensity changes and low frequency variations in the residuals which have to be removed in order to make the very low amplitude frequencies in question detectable. This ``cleaning" was achieved via iterative spline interpolation anchored on the median flux in 2--3 day intervals (in other words applying a high--pass filter) by 3--5\,$\sigma$ clipping to remove outliers and repeating this procedure 20 times. As a result of this rather arbitrary procedure any signal is gradually suppressed towards low frequencies.

The resulting \smei\ time series of \hd\ is homogeneously sampled without significant aliasing. The median cadence of about 101.6\,min results in a Nyquist frequency of $\sim$7\d , which is high enough to cover the intrinsic variability of \hd . The Fourier amplitude spectrum of the \smei\ data set is shown in Fig.\,\ref{fig:smei}. Compared to the BTr spectrum, the formal frequency resolution is more than 18 times better ($\sim$0.00035\d compared to 0.0064\d\ for the BTr data), while the noise level in the Fourier domain is obviously much higher ($\sim$0.53\,ppt compared to 0.11\,ppt in the BTr data). The photometric precision is, however, sufficient to clearly detect the three largest-amplitude doublets $\nu_6$, $\nu_7$, and $\nu_9$ found in the BTr observations. In fact, the doublets turn out to be triplets with the central component having a smaller (or comparable) amplitude than the wing components. This clearly indicates that the BTr observations are not long enough to resolve all three components of the intrinsic triplets even when using our present frequency analysis method. The frequencies (and even the amplitudes) of the wing components (red vertical lines in the top panels of Fig.\,\ref{fig:smei}) agree well, however, with the signal found in the \smei\ spectrum. Also the central components, which were estimated from the midpoints of the (BTr) wing components, are in good agreement.

To quantify the agreement we tried to extract the individual frequencies from the \smei\ observations but find a classical pre-whitening sequence to be insufficient as the individual frequencies split up in many components indicating strong amplitude modulations. This is already visible in the actual spectrum showing multiple side-lobes around the expected positions of the frequencies. Such a structure reminds of the pattern of intrinsically damped and stochastically excited modes (so-called solar-like oscillations) produce in the Fourier spectrum. This is why we fit a sequence of three Lorentzian triplets to the spectrum,
\begin{equation} \label{eq:lor}
P(f) = P_0 + \sum_n {\sum_{m=-1}^1{\frac{h_n \zeta_m(i)}{1 + 4\,[\pi \tau \,(f - f_n - m \cdot \delta f_{rot,n})]^2}}},
\end{equation}    
where $P$ is the Fourier power, $f_n$ is the frequency of the central component of the $n$-th triplet and $\delta f_{rot,n}$ is its rotational splitting. For simplicity we use a single lifetime parameter $\tau$ for all modes. The mode height $h_n$ scales with a geometric factor $\zeta_m(i)$, which depends on $m$ and the inclination $i$ at which the pulsation axis is seen. According to \cite{gizon2003} this geometric factor is $\cos^2{i}$ for the $m=0$ component and $0.5\sin^2{i}$ for the $|m|=1$ components of a dipole mode. More details about the detection of Lorentzian profiles are given by, e.g., \cite{Gruberbauer2009}. We again use \textsc{MultiNest} for the fit and find a best-fit inclination and mode lifetime of 68$\pm$5\degr\ and 680$\pm$110\,d, respectively. The central mode frequencies and rotational splittings are listed in Tab.\,\ref{tab:split} and the best-fit sequence of Lorentzian triplets is shown in the middle panels of Fig.\,\ref{fig:smei}. The triplets at about 0.88 and 0.98\d\ are well defined so that we can test the assumption of symmetric splittings for them. A fit with a modified Eq.\,\ref{eq:lor} (where we allow for individual splittings) indeed shows that the wing components are equally separated from the central component within the uncertainties of about 5\%. 

We emphasise here that it does not necessarily mean that modes have indeed a stochastic nature, if Lorentzians work well in extracting the mode frequencies from the \smei\ observations. In fact, this is very difficult to prove, because one has to show that the signal phase is not coherent, which requires continuous observations that cover many lifetime cycles of the mode. The \smei\ data cover about five lifetime cycles (if the modes were stochastic), which is likely not enough to verify a stochastic nature. We do, however, find a strong amplitude modulation for the individual frequencies. To quantify this we fit a sine function to a 720\,d long subset of the time series, where we fix the frequency to the value determined earlier by the Lorentzian fit. Moving the subset across the \smei\ data in steps of 100\,d gives the amplitude (and phase) of a given frequency as a function of time. This is shown in the bottom panel of Fig.\,\ref{fig:smei}. We tested various window sizes but always find the same general behaviour. The strongest modulation is found for the triplet at about 0.84\d . While the amplitude of the central component is fairly constant, the wing components vary in amplitude by a factor of more than four. Also interesting is that the variations are periodic with a timescale of about 1\,500\,d and that they are in anti phase. The amplitude minimum of the $m=-1$ component coincides approximately in time with the largest amplitude of the $m=+1$ component, and vice versa. A similar but less pronounced behaviour can be found for the triplet at about 0.98\d . Only the triplet at 0.88\d\ is different. The timescale of its modulation is much longer and the wing components vary approximately in phase. The physical origin for this phenomenon is unknown to us but we note that something similar is seen in the \textit{Kepler} observations of KIC\,10526294 \citep{papics2014}. Even though the authors did not follow up on this, many modes illustrated in their Fig.\,9 show a multiple peak structure typical for amplitude modulation. 

Apart from the three triplets shown in Fig.\,\ref{fig:smei} and a single sharp peak at 0.40013\d\ (for which we have no explanation at this point) we do not find any further significant variability in the \smei\ data. The data set does, however, allow us to verify several assumptions made during the analysis of the BTr observations:
\begin{itemize}
	\item Our interpretation of the nine pairs of close frequencies in the BTr data as symmetrically split doublets is confirmed by the independent \smei\ observations. Even though we can only verify the three largest-amplitude multiplets, the remaining doublets follow the same statistical criteria and only their amplitudes are smaller, but still significant in the BTr time series.
	\item The frequencies of the central components and rotational splittings agree on average within $\sim$1.8$\,\sigma$ and 0.8$\,\sigma$, respectively, between BTr and \smei .
	\item The interpretation of individual frequencies extracted from the BTr observations as independent oscillations requires approximately stable signal amplitudes. Even though we definitely find amplitude modulations, their timescales are long enough to consider the oscillations in first approximation to be stable during the 156\,d long BTr observations. Even for shorter lengths of the subsets, which are used for the bottom panel of Fig.\,\ref{fig:smei}, we do not find evidence for significant amplitude modulations shorter than those mentioned above.
	\item The amplitude modulations also provide a reasonable explanation for the missing modes (Sec.\,\ref{sec:periodspacing}) as it might well be that their amplitudes were below the detection threshold of the BTr observations.
\end{itemize}

Finally we note that we cannot straightforwardly constrain the inclination angle of \hd\ since the observed frequencies are heat-driven modes. In this case, the $2l+1$ components of a rotationally split non-radial mode are not excited to the same amplitude, contrary to solar-like oscillators \citep[e.g.][]{gizon2003}. However, the formally best-fit value of 68$\pm$5\degr gives an amplitude ratio between the central and wing components which is at least not inconsistent with the observed ones. It seems to be plausible that we see the star more equator-on than pole-on.

\section{Asteroseismic analysis} \label{Sec:Seis}

To interpret the observed rotational splittings in terms of internal differential rotation we need a representative stellar model for \hd . We therefore compare the observed g modes to non-adiabatic pulsation modes computed with the GYRE stellar oscillation code \citep{townsend2013} for a grid of non-rotating equilibrium stellar models along stellar evolutionary tracks that pass through the spectroscopic error box (see Tab.\,\ref{tab:SpecParam} and Sec.\,\ref{sec:intro}). The models are calculated with the MESA stellar structure and evolution code \citep{paxton2011,paxton2013}. As we are for the time being only interested in a representative model we restrict the models to a single initial chemical composition of (Y, Z) = (0.28, 0.02) and turn off convective core overshooting. We evolve a set of zero-age main-sequence models with masses ranging from 2.9 to 3.225\,M\sun\ (with steps of 0.025\,M\sun ) until their core hydrogen mass fraction drops below $X_c$ = 0.3. To achieve sufficient resolution along the tracks we limit the evolutionary time steps to 0.2\,Myr, which results in about 5\,200 models with a typical resolution in $X_c$ of 0.002 close to the ZAMS to 0.004 for the most evolved models. 

We then compute $l$ = 1 modes with non-adiabatic frequencies ranging from 0.3 to 1.2\d . To search for a best-fit model we compare the observed frequencies ($\nu_{obs}$) to the theoretical ones ($\nu_{model}$) by computing the reduced $\chi^2$ value \citep[e.g.][]{Pamyatnykh1998,guenther2004},
\begin{equation}
\chi^2 = \frac{1}{N} \sum_{i=1}^N \frac{(\nu_{i,obs} - \nu_{i,model})^2}{\sigma_{i,obs}^2 + \sigma_{model}^2},
\end{equation}
where $N$ and $\sigma_{obs}$ are the total number of observed modes and their frequency uncertainties. The typical numerical uncertainty of the model frequencies ($\sigma_{model}$) is estimated from following the frequency of a specific mode (i.e., with a given radial order) during stellar evolution. We thereby assume the actual numerical uncertainty to be of the order of the point-to-point scatter after subtracting a running average. We find a value of about 0.00003\d , which is comparable to $\sigma_{obs}$ in some cases and therefore not negligible.

%--------------------------------------------------------------------
\begin{figure}[t]
	\begin{center}
	\includegraphics[width=0.5\textwidth]{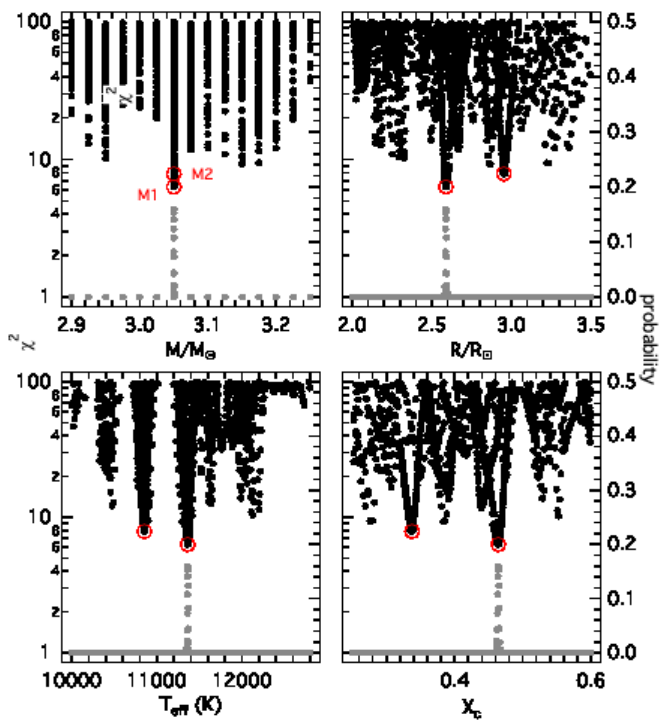}
	\caption{Visualisation of the multi-dimensional $\chi^2$ (left axis; black symbols) and probability (right axis; grey symbols) space that results from the comparsion of the observed and computed MESA/GYRE modes. Red circles mark the two best-fit models.} 
	\label{fig:chisq} 
	\end{center} 
\end{figure}
%--------------------------------------------------------------------
%--------------------------------------------------------------------
\begin{figure}[t]
	\begin{center}
	\includegraphics[width=0.5\textwidth]{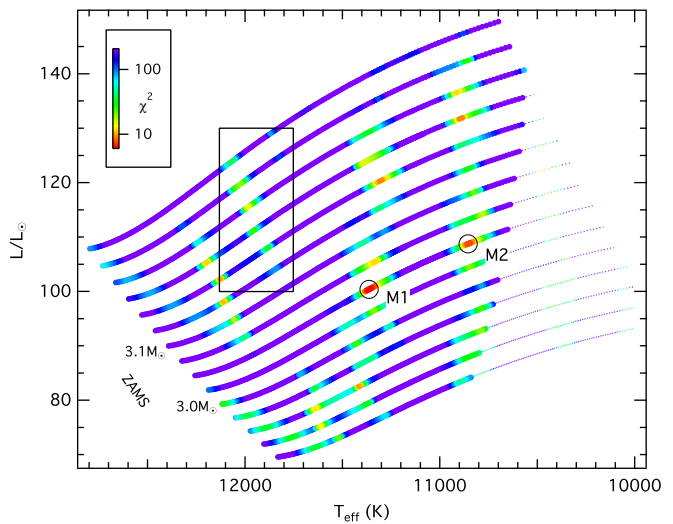}
	\caption{HR diagram showing the model grid that is used to search for a representative stellar model of \hd . The models are color-coded according to their $\chi^2$ value and shown with thick lines when at least one mode excited (i.e., with a positive work integral) and with thin lines when all modes are damped. The two best-fit models are marked with black circles. The black rectangle  indicates the spectroscopic error box (see Tab.\,\ref{tab:SpecParam} and Sec.\,\ref{sec:intro}).} 
	\label{fig:HRD} 
	\end{center} 
\end{figure}
%--------------------------------------------------------------------

The resulting multi-dimensional $\chi^2$ space is shown in Fig.\,\ref{fig:chisq} and illustrates that while the grid resolution along the evolutionary tracks is sufficient (indicated by the smooth decrease and increase of $\chi^2$ around the best-fit model when plotted, e.g., as a function of $R$) the resolution in mass is still too low to find a model whose frequencies agree within the observational uncertainties. We can, however, identify a model that fulfils our requirements (approximately representing the internal structure of \hd). The best-fit model has a $\chi^2$ value of about 6.3, which means that its frequencies matches the observed ones on average within 2.5 (i.e., square root of 6.3) times the average observational errors. The best-fit model parameters are listed in Tab.\,\ref{tab:model} and its frequencies and period spacings are compared to the observational values in Fig.\,\ref{fig:freq} and \ref{fig:Spacings}, respectively. We further note that increasing the mass resolution would not allow us to improve the situation without simultaneously computing models with various chemical compositions and convective core overshoot parameters. \cite{papics2014} and \cite{Moravveji2015} found strong correlations between the mass, the overshoot parameter, and the chemical composition in their seismic analysis of KIC\,10526294 (a star that is very similar to \hd\ in mass and chemical composition, but less evolved), from which we can estimate that turning on core overshooting could potentially increase the mass of the best-fit model by up to 0.1\,M\sun. They do, however, also find that models with a low overshoot parameter fits the observations best. Also the fact that our best-fit model is outside the expected range from spectroscopy in the HR diagram (see Fig.\,\ref{fig:HRD}) is not really troubling since changing the chemical composition would again slightly change the mass and therefore the position in the HR diagram.

%________________________________________________________________Tab. Modelfit
\begin{table}[b]
\begin{small}
\begin{center}
\caption{Properties of the two best fitting models. See text for details. \teff\ is given in K, $M$, $L$, and $R$ in solar units, and the age in Myr. The hydrogen mass fraction in the core $X_c$ is given in units of 1.
\label{tab:model}}
\begin{tabular}{c|cccc|cc}
\hline
\noalign{\smallskip}
& \teff & $M$ & $L$ & $R$ & $\chi^2$ & $p$\\
& \lg   & $X_c$ & age &              & &\\
\noalign{\smallskip}
\hline
\noalign{\smallskip}
M1	& 11363	& 3.05	& 100.4	& 2.589	& 6.30 & 0.15\\
	& 4.096	& 0.464	& 145	&		& &\\
M2	& 10854	& 3.05	& 108.7	& 2.953	& 7.85 & 1.5e-4\\
	& 3.982	& 0.338	& 199	& 		& &\\
\noalign{\smallskip}
\hline
\end{tabular}
\end{center}
\end{small}
\end{table}
%__________________________________________________________________

A disadvantage of the $\chi^2$ method is its inability to provide uncertainties and therefore to set a limit on which models (within the grid) represent the observations and which do not. The ``second-best''-fit model M2 (with its position in the HR diagram significantly different from M1) has a $\chi^2$ of about 7.85 and therefore fits the observed frequencies marginally less well. 
%From a probabilistic point of view this does, however, not exclude the model. 
We therefore follow the approach of \cite{kallinger2010} to determine the Bayesian model probability and find that for our model grid about 99.9\% of the total probability is concentrated in the close vicinity (about $\pm$15\,K) of the best-fit model M1 (M1 itself has a $p$ of about 0.15) and that there is only a marginal probability that M2 provides the best representation of the observations (see Tab.\,\ref{tab:model}). Even though this would already rule out M2 we further investigate it in order to check if a slightly different internal structure (M2 is more evolved than M1 and has a smaller relative core size) affects the further analysis.

%--------------------------------------------------------------------
\begin{figure}[t]
	\begin{center}
	\includegraphics[width=0.5\textwidth]{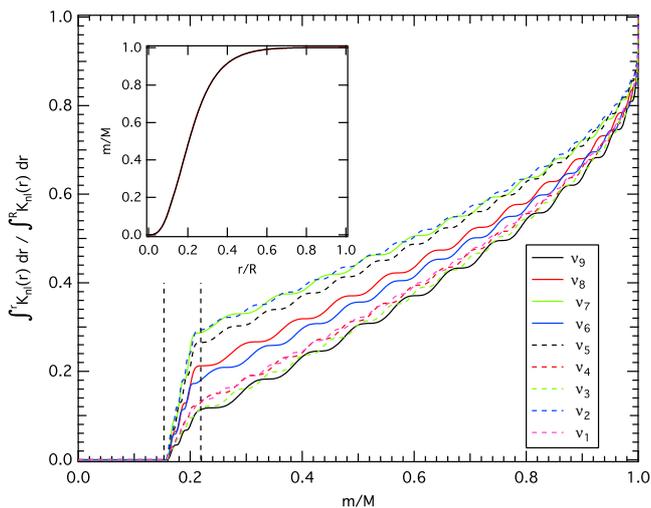}
	\caption{Normalised integrated rotational kernels for the nine dipole modes in model M1 representing best the observed frequencies in \hd . The insert shows how the mass is distributed in the model. The vertical dashed lines represent the boundaries of the sharp spike in the Brunt-V\"ais\"al\"a frequency (see Fig.\,\ref{fig:BV}) with the inner one marking the boundary of the convective core.}
	\label{fig:kernel} 
	\end{center} 
\end{figure}
%--------------------------------------------------------------------

%--------------------------------------------------------------------
\begin{figure}[b]
	\begin{center}
	\includegraphics[width=0.5\textwidth]{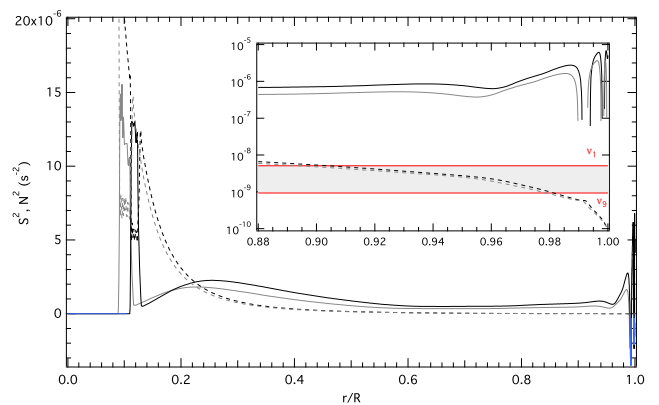}
	\caption{Squared Brunt-V\"ais\"al\"a $N^2$ (solid lines) and Lamb $S^2$ (dashed lines) frequency as a function of the fractional radius for model M1 (black lines) and M2 (grey lines). The blue lines indicate regions where $N^2$ is negative, and are therefore convective. The insert shows the outermost region and the frequency range of the observations (grey-shaded area between the red lines).} 
	\label{fig:BV} 
	\end{center} 
\end{figure}
%--------------------------------------------------------------------

\section{Internal rotation profile}  \label{Sec:Rot}

Building on the measurements of \cite{papics2014} for KIC\,10526294, \citet{triana2015} provided the first internal rotation profile of an unevolved intermediate-mass B-type star. They found the star to rotate near its core-envelope boundary with a period of about 71\,d and while their seismic data point towards a counter-rotating profile within the radiative envelope they cannot rule out rigid rotation. Such results are key to tackle one of the big open questions in stellar evolutionary theory: the transport of angular momentum inside stars. With the nine rotationally split g modes identified in \hd\ we are able to provide another example of an internal rotation profile for a star very similar to KIC\,10526294.

If we assume that the cyclic rotation frequency $\Omega$ depends only on the radial coordinate $r$, then the frequency splitting $\delta \mathrm{f}_{rot}$ of a mode with degree $l$ and radial order $n$ can be written as,
\begin{equation}	\label{eq:splittingKernel}
\delta \mathrm{f}_{rot}(n,l) = \frac{1}{2\pi I_{n,l}} \int_{0}^{R_{*}} K_{n,l}(r) \Omega(r) dr,
\end{equation}
where $I_{n,l}$ is the mode inertia and $K_{n,l}(r)$ gives the unimodular mode kernel, that is a function of the mode's displacement amplitudes \citep[e.g.][]{Cox1980}. $R_{*}$ is the radius of the model.

Normalised integrated versions of the rotation kernels of \hd\ are given in Fig.\,\ref{fig:kernel} and basically show how a rotationally split frequency is accumulated throughout the star. Obviously, different modes are more or less sensitive to rotation in different regions of the star. The mode $\nu_9$, e.g., gains about 10\% of its frequency splitting from rotation in the thin layer above the convective core, where $N^2$ spikes (see Fig.\,\ref{fig:BV}). The mode $\nu_7$, on the other hand accumulates almost three times more of its rotational splitting in the same region. This ``differential'' sensitivity to different parts of the star allow us to resolve the stellar rotation profile. 

From the fact that the modes $\nu_7$ and $\nu_9$ have practically the same observed rotational splitting we can already conclude that \hd\  is either a rigid rotator or it rotates much faster in the outer layers than the inner ones (because otherwise the split frequencies of $\nu_7$ and $\nu_9$ would be different). 
%For a more detailed picture, Eq.\,\ref{eq:splittingKernel} needs to be inverted. 
Several inversion techniques have been developed in the past aiming to determine the internal rotation profile of the Sun. The inversion of Eq.\,\ref{eq:splittingKernel} is, however, a highly ill-conditioned problem that requires, e.g., numerical regularisation. \cite{Beck2014} found that classical approaches like the RLS method \citep[e.g.][]{Dalsgaard1990} or the SOLA technique \citep[e.g.][]{Schou1998} are not well-suited for stars with only a few observed rotational splittings (like red giant stars, but also SPB stars) and easily become numerically unstable or it is very difficult to evaluate the accuracy and especially the reliability of the result. 

\subsection{Forward-modelling of the rotation profile}

We therefore follow the forward modelling approach developed by one of us \citep[TK in][]{Beck2014}. The algorithm computes synthetic rotational splittings for a parameterised rotation profile and compares them to the observed splittings. The form of the profile is thereby very flexible. One can, e.g., implement a linear piece-wise model or a functional form (like a Gaussian or a polynomial). The profile parameters are again fitted with the Bayesian nested sampling algorithm \textsc{MultiNest}. These parameters are $\Omega_n$ in the case of a zonal model with $n$ zones, or the coefficients of a chosen function. The advantage of this approach is again to provide realistic uncertainties and its capability to compare different models and rate which one best represents the observations (e.g., differential rotation vs. rigid rotation). The algorithm was developed to handle the few available rotational splittings in red giants and has proven to give reliable results that are fully consistent with other methods \citep[e.g.][Beck et al., in prep.]{Beck2014}.

The modes that are accessible to our analysis probe the radiative envelope of \hd\ from the boundary of the convective core (at  a radius and mass fraction of $\sim$0.11 and 0.15, respectively) up to about $r = 0.98R_{*}$ ($m>0.9999M_{*}$), above which all frequencies fall above the Lamb frequency (see Fig.\,\ref{fig:BV}), which sets the upper limit for g modes to propagate. We can therefore not expect to directly get information about the core rotation rate and the outer $\sim$2\% of the star. We tried various models for the rotation profile to fit the observed splittings, ranging from rigid rotation, over linear piece-wise models (with up to 9 zones), polynomial functions, Gaussians, multi-Gaussians, to Lorentzian and error functions. A selection of the resulting rotation profiles are shown in Fig.\,\ref{fig:RotProf} and more details are given below: 
 
%--------------------------------------------------------------------
\begin{figure}[t]
	\begin{center}
	\includegraphics[width=0.5\textwidth]{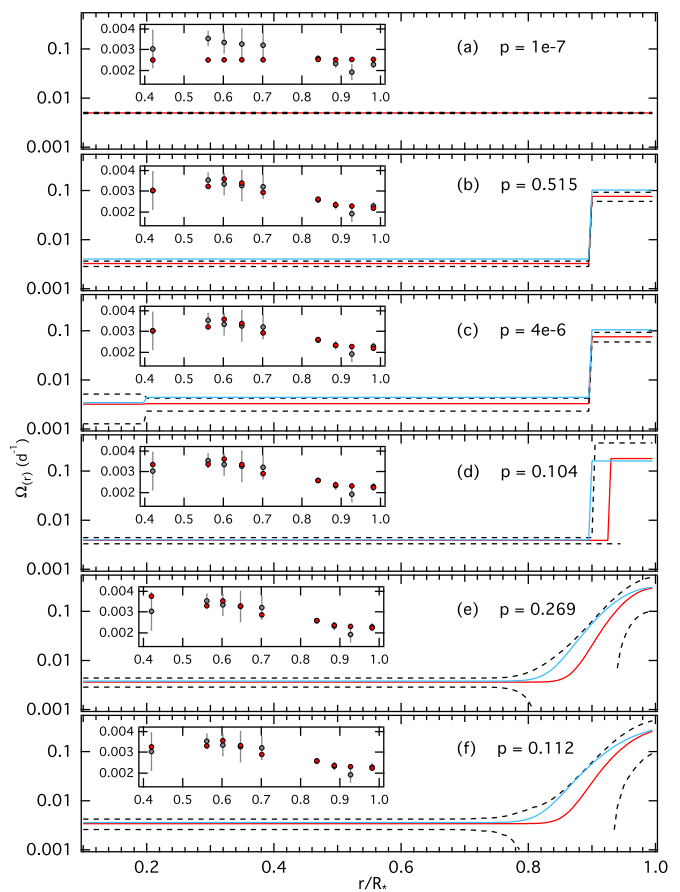}
	\caption{Internal rotation profiles from our forward-modelling approach using various models (panel (\textit{a}): constant rotation; (\textit{b}): two-zone model with fixed zone boundary; (\textit{c}): three-zone model with fixed boundary; (\textit{d}): two zone model with variable zone boundary; (\textit{e}): Gaussian profile centered at the surface; (\textit{f}): error function profile). Red and blue lines indicate rotation profiles resulting from mode kernels of the best-fit models M1 and M2, respectively. The uncertainties (black dashed lines) are only plotted for M1-based profiles for better visibility but are similar for the M2-based profiles. The inserts show the observed (grey symbols) and best-fit (red symbols) rotational splittings as a function of the mode frequency with the axes in \d .} 
	\label{fig:RotProf} 
	\end{center} 
\end{figure}
%--------------------------------------------------------------------

\begin{enumerate}
	\item [(\textit{a})] \textit{Rigid rotation}: Assuming rigid rotation we find a rotation period of 201$\pm$5\,d ($\hat{=}$ 0.0050$\pm$0.0001\d). However, as can be seen in the insert of panel (\textit{a}) in Fig.\,\ref{fig:RotProf}, the synthetic splittings that result from integrating the mode kernels, folded with the rotation profile, do not fit the observed splittings at all. In comparison with other fits the model probability of about 10$^{-7}$ is extremely low. We can therefore rule out an entirely rigidly rotating radiative envelope for \hd .
	
	\item [(\textit{b})] and (\textit{c}) \textit{Multi-zonal rotation profiles}: We test various zonal models with a fixed position of the zone boundary, i.e., ``hard-wired'' in the model and find the best formal model probability (p = 0.515) for a fit with a two-zone model with the zone boundary set at a radius fraction of 0.9 (panel (\textit{b}) in Fig.\,\ref{fig:RotProf}). The inner and outer zones rotate with a period of 314$\pm$43\d ($\hat{=}$ 0.0033$\pm$0.0005\d) and 14$\pm$3\,d ($\hat{=}$ 0.076$\pm$0.018\d), respectively. We do not find another model, even with more zones, that comes close in model probability. As an example we show a three-zone model in panel (\textit{c}) of Fig.\,\ref{fig:RotProf}, where we add an inner zone (0 - 0.2$R_{*}$) but find it to rotate with practically the same rotation rate as the middle zone. This indicates that the data do not support differential rotation in the radiative envelope of \hd\ below a radius fraction of about 0.9. The very low model probability compared to model (\textit{b}) results from the additional parameter in the fit\footnote{Roughly spoken, in a Bayesian concept a model gets assigned a penalty for its complexity so that a model with $n+1$ parameters has to fit the data significantly better than a model with $n$ parameters to get assigned the same (or even higher) model evidence. More details are provided by \citet{jeffreys98}.}.
	
	\item [(\textit{d})] \textit{Two-zone profile with variable zone boundary}: In a next step we tried to locate the transition region between the inner slow and outer fast rotating zone. In contrast to the fixed location of the zone boundary in the original two-zone model we now leave it as a free parameter in the fit. The result is shown in panel (\textit{d}) locating the zone boundary at a radius fraction of $r/R_{*}$ = 0.93$\pm$0.02. While the rotation period of the inner zone remains almost the same, the outer zone now rotates with a period of about 5.5\,d ($\hat{=}$ 0.182\d) significantly faster than for the original two-zone model. However, due to the rotational splittings containing less and less information about rotation when approaching the surface, the uncertainties start to dramatically increase.
	
	\item [(\textit{e})] \textit{Gaussian rotation profile}: So far, the analysis points towards a slowly and rigidly rotating zone that contains almost the entire mass of the radiative envelope, topped by a thin and significantly more rapidly rotating surface layer. The assumption of a sudden increase in rotation speed (by a factor of 20 or so) at the zone boundary seems, however, physically not very plausible. We therefore use a Gaussian rotation profile in which it turns out that we get the best results (i.e., the best model probability) when fixing the centre of the Gaussian to the stellar surface. While the inner rotation period of 292$\pm$76\,d ($\hat{=}$ 0.0034$^{+0.0012}_{-0.0007}$\d) agrees well with the two-zone models, the surface rotation rate is 0.30$\pm$0.21\d\ ($\hat{=}$ 3.33$^{+7.78}_{-1.37}$\,d) -- larger than those of the two-zone models. Since the latter do, however, only give average rotation rates within the zones, we consider the models as equivalent. Also the width of the Gaussian (it drops to half its maximum at $r/R_{*}$ = 0.96$\pm$0.01) is consistent with the zone boundary of the variable two-zone model. From a statistical point of view the probability contrast between the two-zone models and the Gaussian (about 1:2 for model e:b and 5:2 for model e:d) is not enough to prefer any of them. 
	
	\item [(\textit{f})] \textit{Error function rotation profile}: Even though more physical than a two-zone model, the Gaussian model is not as capable of reproducing the observed rotational splittings, indicating that the transition from the rapidly rotating surface to the slowly rotating interior might not be Gaussian. We therefore fit an error function $\Omega(r) = \Omega_0 + 0.5\Omega_{1}(1 + \mathrm{erf}([r-r_{1}]/k))$ to the observed splittings, where $r_1$ gives the location where $\Omega$ has risen to half its maximum value ($\Omega_1$) and $k$ controls the slope of the increase. The advantage of this function is that it can reproduce very steep to shallow transitions from slow to fast rotation. Interestingly, the fit gives the best-fit parameters \makebox{$\Omega_1$ = 0.26$\pm$0.17\d ($\hat{=}$ 3.85$^{+7.26}_{-1.52}$\,d)}, $r_1$ = 0.97$\pm$0.02, and $k$ = 0.06$\pm$0.03, which results in a rotation profile that is very similar to the Gaussian profile. Even though the error function profile is more complex than the Gaussian (it has 4 instead of 3 free parameters) the model probability is 0.112 -- high enough to be considered a reasonable representation of the data.  
\end{enumerate}
Note that we limit the rotation rates to values between $-1$ and 1\d\ for all our fits, hence allowing for counter-rotation \citep[as indicated by][for KIC\,10526294]{triana2015}. However, we did not find a single model that includes a statistically significant counter-rotating zone in \hd .

The basic result of the rotation analysis is that the measured rotational splittings are inconsistent with the entire radiative envelope rotating at a constant rate, providing instead strong evidence (in a statistical sense) for a slowly and rigidly rotating envelope topped by a thin and significantly more rapidly rotating surface layer. There is some weak evidence that this layer reaches about 4\% down in radius ($<0.001$\% in mass) and that the transition between the slow and rapidly rotating zone is Gaussian-like. It is interesting to mention that the short rotation period is compatible -- within the large uncertainties -- with the orbital period of the innermost companion of \hd , suggesting angular momentum transfer from the companion star to the outer envelope of \hd , causing its acceleration.

An important question is how these results depend on the mode kernels used and therefore the specific best-fit stellar model. We repeated the entire rotation analysis with model M2 and find no significant difference to our original analysis with model M1. The only difference is that the rapidly rotating outer layer seems to reach further down (see panels (\textit{d}) to (\textit{f}) in Fig.\,\ref{fig:RotProf}), but the deviations are still within the uncertainties. 

\begin{figure}[t]
	\begin{center}
	\includegraphics[width=0.5\textwidth]{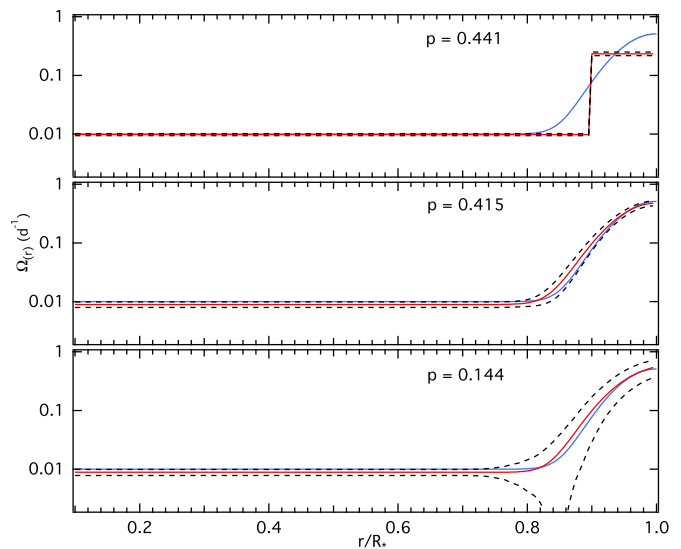}
	\caption{Same as Fig.\,\ref{fig:RotProf} but for synthetic rotational splittings that are computed from a Gaussian rotation profile (blue line). Panels from top to bottom: two-zone model with fixed boundary at $r/R_{*}$ = 0.9, Gaussian profile, and an error function profile.} 
	\label{fig:RotProfSim} 
	\end{center} 
\end{figure}

Finally we test the reliability of our approach, i.e., how well can we reconstruct the internal rotation profile? We therefore compute synthetic rotational splittings based on the M1 mode kernels and arbitrarily assumed rotation profiles (ranging from solid-body rotation, multi-zonal profile, the Gaussians with the centre set to the core and to envelope) and add noise to them. As an example, we show in Fig.\,\ref{fig:RotProfSim} the result of our forward-modelling approach for a simulated Gaussian rotation profile, similar to what we find for \hd . The simulation demonstrates that we can indeed reconstruct the input rotation profile within the limits of the chosen model. This cannot, however, be directly compared to the real data. In the case of the simulation we know the functional form of the profile and therefore ``only'' reconstruct its parameters. For the real data, this is not the case and the relatively large uncertainties (compare panels (\textit{e}) and (\textit{f}) in Fig.\,\ref{fig:RotProf} to the respective profiles in Fig.\,\ref{fig:RotProfSim}) do also reflect that the model does not perfectly match the functional form of the real rotation profile. The data do not, however, allow one to better constrain the real form of the rotation profile without over-fitting.

%--------------------------------------------------------------------
%--------------------------------------------------------------------

\section{Multiplicity of \hd}

\hd\ has been known for almost a century to be a gravitationally bound multiple system. \cite{young1921} measured the radial velocity (RV) of the system 54 times during 1920/21 and found the RV's to be consistent with a 3.3137\,d circular binary orbit. The residuals to the fit were, however, much larger than the uncertainties of the measurements, which is why the system was found to be ``peculiar''. \cite{guthnik1938,guthnik1939,guthnik1942} picked up on the peculiarity and improved the original orbital elements of Young (based on 90 new measurements between 1936 and 1940) but also had no explanation for the persistent large residuals. He also detected photometric variability that was not compatible with the binary orbit. Two Cepheid-like frequencies, 0.838026\d and 0.885645\d, which are consistent with $f_1$ and $f_4$ of the BTr observations (see Tab.\,\ref{tab:freq}), were found in these old photoelectric data, interrupted by periods of irregular variations or constant brightness. \cite{Hoffleit1977} noted that the difference between these two periods is exactly 1/21\,d but had no interpretation for this. In fact, 1/21\,d corresponds to about 4100\,s and is therefore a reasonable estimate for the g-mode period spacing in \hd. 

\cite{gieseking1978} obtained 60 new RV measurements in 1975-77 and improved the original orbital period to 3.3131168$\pm$0.000008\,d using the combined data set, which was finally large enough to discover the second companion with a period of 154.09$\pm$0.02\,d, assuming circular orbits. \cite{barlow1989} re-reduced the observations of \cite{young1921} and used all available data to find a new solution for the triple system with periods of 3.31311718$\pm$0.0000045 and 154.072$\pm$0.018\,d, and also found that the long-period (LP) companion is in a non-circular orbit with an eccentricity of 0.311$\pm$0.057.

\subsection{New spectroscopy of \hd }
Motivated by the fact that the earliest RV measurements date back almost a century and that no high-resolution spectrum of \hd\ was available for a spectroscopic analysis we organised new spectroscopic observations. A total of 26 spectra (see Tab.\,\ref{tab:ObsLog}) were obtained in May and August 2016 with the \Hermes spectrograph \citep{Raskin2011,RaskinPhD}, mounted on the 1.2\,m \textsc{Mercator} telescope on La Palma, Canary Islands, Spain.  The \Hermes spectra cover a wavelength range between 375 and 900\,nm with a spectral resolving power of R $\simeq$ 85\,000. The wavelength reference was obtained from emission spectra of thorium-argon-neon reference frames in close proximity to the individual exposures. The extraction and the data reduction of the observed stellar spectra were performed with the instrument-specific pipeline \citep{Raskin2011}. Additionally we obtained a single spectrum in late April 2016 with the echelle spectrograph NES (R $\simeq$ 43\,000) mounted to the 6\,m SAO telescope of the Russian Academy of Sciences.

%--------------------------------------------------------------------
\begin{figure}[t]
	\begin{center}
	\includegraphics[width=0.5\textwidth]{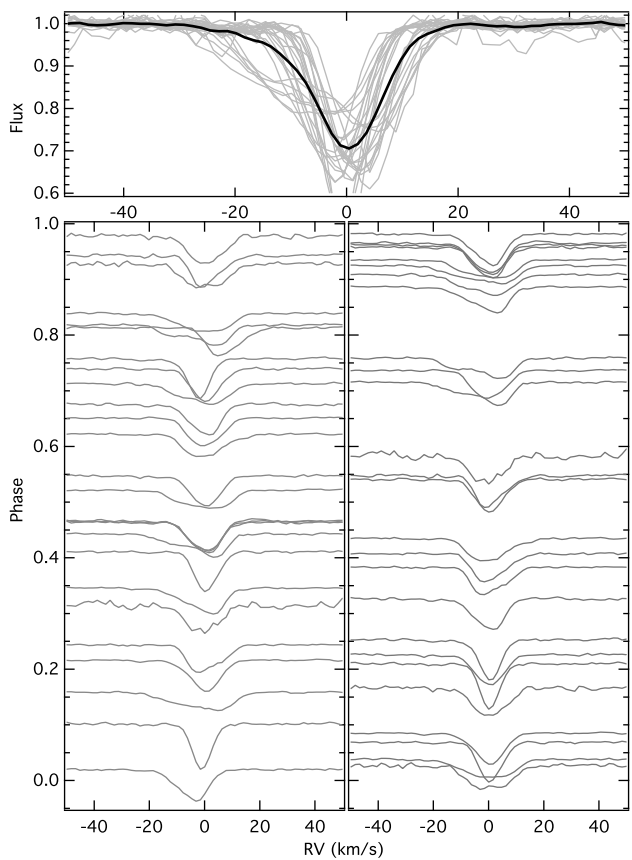}
	\caption{Line profile of the \ion{Fe}{II} (4233.1\AA) line. The top panel shows the individual (grey lines) and average (black line) line profiles corrected for the orbital motion. The bottom panels show the individual profiles arbitrarily scaled and vertically shifted according to their orbital (left) and pulsation (right) phase ($f_1$ from Tab.\,\ref{tab:freq}).} 
	\label{fig:lineprofiles} 
	\end{center} 
\end{figure}
%--------------------------------------------------------------------

The radial velocities were extracted by cross-correlating the individual spectra with a template spectrum and fitting a Gaussian to the response function. The RV measurements include the combined orbital motions as well as a component from the surface oscillations of \hd\ (see Fig.\,\ref{fig:lineprofiles}). As the latter will influence our orbital solution we have to correct for them. In a first step we increased the formal estimates of the uncertainties by a factor of five ($\bar\sigma_{RV} = 0.13 \rightarrow 0.65$\,km/s) to make them compatible with the scatter of the first three RV values, which were obtained within a few minutes with \Hermes at \textsc{Mercator} and which should practically have the same orbital velocity. The resulting RV uncertainties are still much smaller than the typical uncertainties of the old measurements of about 2.8\,km/s. The new data should therefore help to significantly improve the orbital elements of the triple system and allow us to search for changes in the orbital periods of the system.

\subsection{Line profile variations}

A closer look at the individual spectra reveals that the line profiles are in many cases asymmetric and that the profile of a given spectral line changes from spectrum to spectrum. This can be due to the spectral signature of an unresolved companion, but -- in our case -- is more likely due to non-radial oscillations in \hd, which can be best tested with a strong and unblended spectral line. Even though \hd\ is a relatively hot star, such lines are rare. A good candidate is the \ion{Fe}{II} line at 4233.1\,\AA. In Fig.\,\ref{fig:lineprofiles} we plot the individual spectra corrected for the orbital motions and vertically shifted according to orbital phase (bottom left) and pulsation phase (bottom right) with $P_\mathrm{SP}$ from Tab.\,\ref{tab:Borbit} and $f_1$ from Tab.\,\ref{tab:freq}, respectively. If the line profile variations are due to unresolved spectral lines of  the companion star the asymmetries in the spectra would be in phase and the ``motions'' across the line profile would show some structure. This is clearly not the case. If we phase-fold the spectra with the period of the strongest photometric frequency (1.19217\,d; Tab.\,\ref{tab:freq}), the asymmetries are now much more in phase. For example, the line profiles of the two spectra at a phase of about 0.55 are almost identical even though they were obtained almost 6 days apart. Furthermore, the line profile variations show some systematics, e.g, a ``bump'' at a phase of about 0.4 moving towards higher RV half a period later (i.e., at a phase of 0.9). Unfortunately, the spectra are too noisy and the phase coverage is not complete enough for a detailed pulsation mode identification. The available spectra do, however, indicate that the line profile variations are due to non-radial oscillations and that their morphology is not inconsistent with $l=1$ modes \citep[e.g.][]{aerts2010}.

%________________________________________________________________Tab. ObsLog
\begin{table}[t]
\begin{small}
\begin{center}
\caption{Journal for the SAO (first line) and \textsc{Mercator} observations with the original and pulsation-corrected radial velocity measurements (RV and RV$_\mathrm{cor}$, respectively). The listed uncertainties are multiplied by a factor of five to account for the line profile variations. 
\label{tab:ObsLog}}
\begin{tabular}{cccrrr}
\hline
\noalign{\smallskip}
%&\multicolumn{2}{c|}{BTr} &\multicolumn{2}{c|}{BLb} &&\\
BJD - 2\,420\,000		&	ExpTime		&S/N	&	RV	&RV$_\mathrm{cor}$	&$\sigma_\mathrm{RV}$\\
(d)	& (s)	& &	\multicolumn{3}{c}{(km/s)}\\
\noalign{\smallskip}
\hline
\noalign{\smallskip}
37\,506.54653 &1200	&50		&$-26.80$ &-   &0.30\\	
\noalign{\smallskip}
37\,521.71877	&300	&96		&$-25.04$ & $-$27.00  &  0.38	\\
37\,521.72340	&400	&117	&$-24.68$ & $-$26.53  &  0.33	\\
37\,521.72803	&300	&106	&$-24.39$ & $-$26.13   &  0.39	\\
37\,525.73500	&120	&70		&$-11.88$ & $-$7.60   &  0.32	\\
37\,526.56969	&130	&63		&$-28.05$ &  $-$27.55  &  0.63	\\
37\,526.73557	&110	&60		&$-38.95$ & $-$35.07   &  0.46	\\
\noalign{\smallskip}
37\,605.52232	&450	&111	& $-4.93$ & $-$5.68   &  0.24	\\
37\,605.70650	&500	&124	&  0.01 & $-$4.21   &  0.64	\\
37\,606.39134	&600	&123	&$-36.07$ & $-$31.43   &  0.38	\\
37\,606.66337	&450	&109	&$-40.15$ & $-$39.78   &  0.20	\\
37\,607.46975	&450	&98		&$-38.04$ &  $-$34.02  &  0.97	\\
37\,607.68725	&450	&115	&$-25.31$ & $-$21.55   &  0.26	\\
37\,608.48371	&450	&115	& $-0.85$  & $-$0.25  &  0.39	\\
37\,608.68911	&450	&131	& $-1.29$ & 3.07   &  1.02	\\
37\,609.45007	&600	&106	&$-20.29$ & $-$24.30   &  0.69	\\
37\,610.44661	&900	&114	&$-41.82$ & $-$45.95   &  0.79	\\
37\,610.68638	&248	&31		&$-32.51$ & $-$35.77   &  0.86	\\
37\,611.45103	&400	&126	&$-12.79$ & $-$13.38   &  0.25	\\
37\,611.70037	&450	&122	&6.29 &1.68	&  0.76	\\
37\,612.41983	&450	&134	&$-19.94$ &  $-$15.96  &  0.60	\\
37\,613.47615	&400	&125	&$-52.73$ & $-$48.13   &  1.28	\\
37\,613.66746	&400	&132	&$-47.51$ & $-$44.48   &  0.23	\\
37\,614.41879	&400	&128	&$-28.29$ & $-$27.70   &  1.64	\\
37\,614.68173	&300	&113	&$-19.71$ & $-$15.07   &  1.02	\\
37\,615.40370	&600	&107	& $-3.51$ & $-$7.27   &  0.19	\\
37\,615.66289	&600	&124	&$-13.84$  & $-$11.90  &  1.87	\\
\noalign{\smallskip}
\hline
\end{tabular}
\end{center}
\end{small}
\end{table}
%__________________________________________________________________
% mean total line broadening (~vsini) = 7.0452866    +/-   1.8249182 from pipline masking
%f1=[0.842388046,4.64798,0.583941]	;*** for BJD > 30000
%f2=[1.31303366,3.22527218,0.763421]	;*** for BJD < 30000

\subsection{Atmospheric parameters and $v\sin i$} \label{sec:atmo_par}

It is evident that the shape and depth of spectral lines in a single spectrum are strongly affected by  oscillations, impeding the determination of atmospheric parameters and the rotational line broadening. We can, however, minimise this effect by averaging all spectra with the additional benefit of  increasing the SNR and wavelength resolution. We therefore oversample the individual \Hermes spectra, correct them for the orbital motions (original RV in Tab.\,\ref{tab:ObsLog}), and compute a weighted average, where the weights are given by the SNR of the individual spectra. The resulting average \Hermes spectrum has a SNR of at about 700 (at 4700\,\AA).

\begin{table}[!t]
\centering
\caption{Atmospheric parameters of \hd, derived with the \textsc{SME} package, using a mean of the \Hermes spectra. } 
\label{tab:SpecParam}
\begin{small}
\begin{tabular}{lccc}
\hline
\noalign{\smallskip}
 &LLmodels&ATLAS models&Takeda et al.\\
\noalign{\smallskip}
\hline
\noalign{\smallskip}
 \teff\ (K)	&11950$\pm$200  &	12000$\pm$200 &12193$\pm$350  \\
 \lg     & 4.15$\pm$0.07 & 4.13$\pm$0.07 &4.24$\pm$0.2 \\  
Metallicity & 0.02$\pm$0.16 & 0.02$\pm$0.15 & \\ 
 \vs\ (km/s)& 9.8$\pm$2.2  &  &  15    \\ 
\noalign{\smallskip}
\hline
\end{tabular}
\end{small}
\end{table}

The atmospheric parameters of \hd\ were then derived using the \textsc{SME} (Spectroscopy Made Easy) spectral package \citep{1996AAS..118..595V, 2016arXiv160606073P}, designed to perform an analysis of stellar spectra using spectral fitting techniques in the LTE (local thermodynamic equilibrium) approximation. We choose five large spectral regions 4000-4200\,\AA, 4200-4400\,\AA, 4400-4700\,\AA, 4700-5200\,\AA, and 5200-5600\,\AA\ for the fitting procedure. They include three hydrogen lines, H$\delta$, H$\gamma$, and H$\beta$, which are more sensitive to surface gravity than to effective temperature, as well as a large number of lines from light and Fe-peak elements. Few lines of the elements \ion{He}{I}, \ion{Si}{II}, \ion{Mg}{II} that are influenced by NLTE (deviations from the local thermodynamic equilibrium) effects were excluded from the fitting. The atomic parameters for all spectral lines were taken from the third version of the VALD database \citep{2015PhyS...90e4005R}. \textsc{SME} allows one to automatically derive fundamental parameters of a stellar atmosphere: effective temperature, surface gravity, metallicity, radial and projected rotational velocities, micro- and macroturbulent velocities. These parameters are determined by interpolation in a grid of LLmodels \citep{2004AA...428..993S, 2012MNRAS.422.2960T}, which are most suitable for the analysis of A to late B-type stars. \textsc{SME}-derived parameters in BAFGK-stars were checked by a comparison with other spectroscopic or independent (e.g., interferometry) determinations \citep{2015ASPC..494..308R, 2016MNRAS.456.1221R} and were found to agree within 1-2\%\ in effective temperature and $\pm$0.1\,dex in surface gravity. We also run \textsc{SME} with Kurucz's model grid\footnote{http://kurucz.harvard.edu/grids.html}. The final determinations are given in Tab.\,\ref{tab:SpecParam} together with the results of \cite{takeda2014}, which were derived from calibrations of Str\"omgren photometric indices.

%--------------------------------------------------------------------
\begin{figure}[h]
	\begin{center}
	\includegraphics[width=0.48\textwidth]{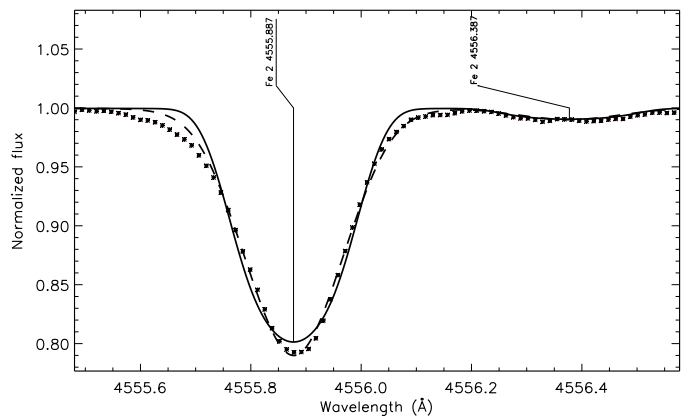}
	\caption{Comparison of observed line profiles (asterisks) with synthetic profiles (full line), which are rotationally broadened with \vs\ = 9.8\,km/s  and with synthetic profiles convolved with \vs\ = 7.9\,km/s and \Vmac = 7.2\,km/s (dashed line).} 
	\label{fig:vsini} 
	\end{center} 
\end{figure}
%--------------------------------------------------------------------

We find that the spectroscopically derived \teff\ and \lg\ values agree well with those derived from photometry. However, the error for the rotational velocity is rather large, obviously due to pulsation. In Fig.\,\ref{fig:vsini} we compare the observed line profiles of a rather strong (4555\,\AA) and weak (4556\,\AA) \ion{Fe}{II} line with synthetic profiles. While the latter, only broadened by rotation, does not reproduce the observations very well, a combination of rotation and macro-turbulence broadening gives a significantly better fit. There is of course no classical macro-turbulence in the atmosphere of a main-sequence B-star; hence, for \hd\ the extra broadening is likely due to pulsation. Such an extra broadening is also observed in the atmospheres of, e.g., $\delta$ Sct stars \citep{Mittermayer2003} and rapidly oscillating Ap stars \citep{2007A&A...473..907R, 2008MNRAS.389..903S} but also for SPB stars \citep{DeCat2002}. 
Following the approach of \cite{Murphy2016} we estimate the pulsational broadening to be of the order of 5.5\,km/s, which reduces the true projected rotational velocity \vs\ to about 8$\pm$2\,km/s.

\subsection{A new solution for the triple system}

With a total of 231 spectra covering slightly more than 96 years we try to derive a new orbital solution for \hd . In a basic Keplerian model for a triple system the primary's radial velocity $RV$ at an epoch $t$ can be computed according to,
\begin{equation} \label{Eq:RV}
RV_{(t)} = \gamma + \sum_{i=1}^2 K_i\, [\cos(\omega_i + \varphi_{(e_i,t,T_i,P_i)}) + e_i ,\cos \omega_i],
\end{equation}
where $\varphi$ is the true anomaly, which depends on $t$, the eccentricity $e$, the epoch of the periastron $T$, and $P$, the orbital period. $\gamma$, $K$, and $\omega$ give the systemic velocity, the semi-amplitude of the visible component, and the longitude of periastron, respectively. For a first fit to the observations we again use \textsc{MultiNest} and find a solution which is consistent with the results of \cite{barlow1989}. 

%--------------------------------------------------------------------
\begin{figure}[t]
	\begin{center}
	\includegraphics[width=0.5\textwidth]{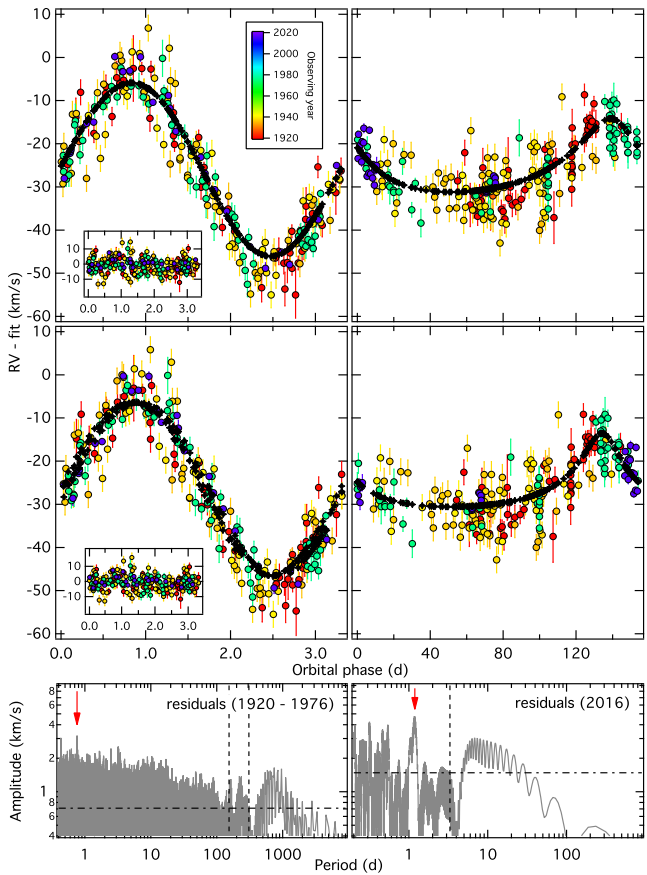}
	\caption{Keplerian fits to the corrected radial velocity measurements for the short-period (left) and long-period (right) companions. The data and fits in the left panels are corrected for the LP orbit and vice versa. The epoch of the observations is colour-coded and black symbols correspond to the best-fit model. The upper and middle panels correspond to fits with a fixed period and a changing period (for the short-period orbit), respectively. The inserts give the residuals. The bottom panels show the Fourier amplitude spectra of the residuals to the initial fit for the 1920-1976 (left) and 2016 (right) data with the dash-dotted horizontal lines indicating the noise level. Vertical dashed lines give the long-period and second harmonic (left) and short-period orbital period (right), respectively. Significant signal is indicated by red arrows.} 
	\label{fig:BinaryOrbit} 
	\end{center} 
\end{figure}
%--------------------------------------------------------------------

We then compute the Fourier power spectrum of the residual velocities (see bottom panels in Fig.\,\ref{fig:BinaryOrbit}). There is significant variability in the residuals, but remarkably different for the old and new data sets. While we find a peak at about 1.3130\d with a SNR of 4.8 in the residuals of the old data, there is essentially no power at this frequency in the 2016 data. Contrary to this, we find the strongest peak at about 0.84\d with a SNR of about 3.7 in the 2016 data, which is not detectable in the old data. While this peak can clearly be attributed to pulsation, we have no explanation for the 1.31\d\ periodicity. We note that also a 1\d\ alias of the SP orbital frequency can be excluded. To improve the orbital solution, we correct for both periodicities according to:
\begin{equation}
RV_{cor} = RV - A \sin \left [2\pi \left (\frac{\mathrm{mBJD}}{\Phi} + \phi \right) \right ]
\end{equation}
where $\{$$A$, $\Phi$, $\phi$$\}$ = $\{$3.22\,km/s, 0.7616\,d, 0.763$\}$ and $\{$4.64\,km/s, 1.1871\,d, 0.584$\}$ for the old (mBJD $<$ +\,24\,000) and new (mBJD $> +\,37\,000$) data, respectively. The corrected radial velocities from 2016 are listed in Tab.\,\ref{tab:ObsLog}.

In a next step we repeat the fit of Eq.\,\ref{Eq:RV} to the corrected radial velocities. As usual the best-fit parameters and their 1$\sigma$ uncertainties are determined from the posterior probability distributions delivered by \textsc{MultiNest}. The resulting orbital elements are listed in Tab.\,\ref{tab:Borbit} and agree with those of the previous fit, but are typically twice as precise as before and are significantly more precise (up to a factor of ten) compared to the solution of \cite{barlow1989}. Most notably, the period of the short-period (SP) orbit is now precise to about 8\,ppm (or 0.034\,s). This might appear unrealistically small but one has to keep in mind that the data cover more than 10\,000 revolutions. While in previous analyses the SP orbit was assumed to be circular we do actually allow also for elliptic orbits, but find the eccentricity to be close to zero ($e_\mathrm{SP} = 0.015\pm0.003$).

%________________________________________________________________Tab. Orbit Parameter
\begin{table}[t]
\begin{small}
\begin{center}
\caption{Orbital solutions for the \hd\ triple system. Given are the systemic velocity ($\gamma$), the epoch of periastron ($T$), the orbital period ($P$), the period change rate ($1/P \cdot dP/dt$), the longitude of the periastron ($\omega$), the orbital eccentricity ($e$), the radial velocity semi amplitude ($K$) of the visible component, the mass function ($f_{(m)}$), and the projected semi-major axis ($a \sin(i)$). For the variable-period fit $P_\mathrm{SP}$ is defined for $T_\mathrm{SP}$ and $rms$ corresponds to the standard deviation of the residuals with the bracketed value computed for the 2016 measurements. The probability is calculated from the Bayesian model evidence ($z$) according to $p = z_n / \sum z$. 
\label{tab:Borbit}}
\begin{tabular}{l|rr}
\hline
\noalign{\smallskip}
%&\multicolumn{2}{c|}{BTr} &\multicolumn{2}{c|}{BLb} &&\\
&constant Period&variable Period\\
\noalign{\smallskip}
\hline
\noalign{\smallskip}
&\multicolumn{2}{c}{short-period (SP) orbit} \\
\noalign{\smallskip}
\hline
\noalign{\smallskip}
$T_\mathrm{SP}$ (d) 		&2\,457\,613.15$\pm$0.08			&2\,457\,613.14$\pm$0.15\\
$P_\mathrm{SP}$ (d) 		&3.3131662$\pm$0.0000004		&3.3131574$\pm$0.0000028\\
$\frac{1}{P} \frac{dP}{dt}$ (yr$^{-1}$)			&-							&3.2$\pm$1.0e-8\\
$\omega_\mathrm{SP}$ (rad)	&2.38$\pm$0.16				&2.36$\pm$0.29\\
$e_\mathrm{SP}$				&0.015$\pm$0.003				&0.015$\pm$0.005\\
$K_\mathrm{SP}$ (km/s)			&20.09$\pm$0.04				&20.07$\pm$0.08\\
$f_{(m)}$ (M\sun)		&0.00278$\pm$0.00002			&0.00277$\pm$0.00003\\
$a_\mathrm{SP,1} \sin(i)$ (R\sun)		&1.316$\pm$0.003				&1.314$\pm$0.005	\\
\noalign{\smallskip}
\hline
\noalign{\smallskip}
&\multicolumn{2}{c}{long-period (LP) orbit} \\
\noalign{\smallskip}
\hline
\noalign{\smallskip}
$T_\mathrm{LP}$ (d) 		&2\,457\,592.7$\pm$0.5			&2\,457\,593.8$\pm$0.7\\
$P_\mathrm{LP}$ (d) 			&154.119$\pm$0.008			&154.151$\pm$0.005\\
$\omega_\mathrm{LP}$ (rad)	&0.27$\pm$0.05				&0.11$\pm$0.04\\
$e_\mathrm{LP}$ (1)			&0.377$\pm$0.026				&0.499$\pm$0.015\\
$K_\mathrm{LP}$ (km/s)			&8.51$\pm$0.10				&8.52$\pm$0.17\\
$f_{(m)}$ (M\sun)			&0.0078$\pm$0.0003			&0.0065$\pm$0.0005\\
$a_\mathrm{LP,1} \sin(i)$ (R\sun)		&25.9$\pm$0.3				&25.9$\pm$0.5	\\
\noalign{\smallskip}
\hline
\noalign{\smallskip}
$\gamma$ (km/s)		&-25.78$\pm$0.11				&-26.05$\pm$0.23\\
rms (km/s) &4.68(2.83)& 4.67(2.82)\\
probability		&0.002							&0.998\\
\noalign{\smallskip}
\hline
\noalign{\smallskip}
\end{tabular}
\end{center}
\end{small}
\end{table}
%__________________________________________________________________

Close binary systems are subject to several physical perturbations, like tidal friction, magnetic braking, mass transfer, etc.   \citep[see, e.g.,][]{Eggleton2006} that increase/decrease the angular momentum of the system and therefore affect the orbital period. In fact a number of close binaries are known that show period changes of the order of some ten milliseconds per year for orbital periods of a few hours \citep[e.g.][]{lohr2012}. The orbital period of the SP companion of \hd\ is much longer, but given the long time base of the observations we might be able to find a period change. We therefore replace the constant orbital period $P_\mathrm{SP}$ in Eq.\,\ref{Eq:RV} by the following expression: 
\begin{equation}
P'_{(t)} = P_\mathrm{SP} \left [1 + \frac{1}{P}\frac{dP}{dt} (T_\mathrm{SP} - t)/365.25 \right ], 
\end{equation}
where $P_\mathrm{SP}$ gives the orbital period at $T_\mathrm{SP}$ and 1/P$ \cdot dP/dt$ is the relative change rate per year. A fit with \textsc{MultiNest} gives a statistically significant period change rate of (3.2$\pm$1.0)$\times$10$^{-8}$. The model evidences of the fits with constant and variable $P_\mathrm{SP}$ have an odds ratio of about 1:500, clearly favouring the variable-period fit. Both fits are shown in Fig.\,\ref{fig:BinaryOrbit} and their orbital elements are listed in Tab.\,\ref{tab:Borbit}. We find only $P_\mathrm{LP}$ and $e_\mathrm{LP}$ to differ by more than 3$\sigma$ between the two models, with the variable-period model resulting in a slightly longer and more eccentric orbit for the long-period companion. Remarkably, the rms scatter of the residuals is still much larger than the uncertainties of the RV measurements and the residuals do not show any systematic trends. We therefore attribute the excess in scatter to unresolved oscillations, but tidal effects might also play a role \citep[e.g.][]{Sybilski2013,zahn2013}.

The determined period change rate indicates that the orbital period of the short-period companion is now about 0.88\,s shorter than in 1921, which is a very small value. But one should keep in mind that the SP period is determined with an accuracy of 0.034\,s! Even though we find strong evidence against a stable period we cannot rule out other explanations, like the influence of another star being gravitationally bound to the system. In fact, there is a known visual companion located about 3.4\,arcsec from the main component. Its spectral type was indicated between A0 and A7 \citep{Meisel1968,Abt1984} with an orbital period likely longer than 1000\,yr. Clarity might be achieved by searching for variability in the $\gamma$-velocity of the triple system. If the detected period change was entirely caused by light-time effects due to orbital motions of an undetected forth component, the expected centre-of-mass velocity change would be on the order of 90\,cm\,s$^{-1}$. Obviously, this is well below the accuracy of the present observations (see bottom panels of Fig.\,\ref{fig:BinaryOrbit}). On the other hand, a 1000 year-long orbit would not be detectable in such a Fourier spectrum, even if we had cm\,s$^{-1}$ precision for our 96 year-long observations. We can, however, search for a trend in the residuals of our analysis. We therefore fit the residuals to a constant period RV time series with first- and second-order polynomials and calculate their \textsc{MultiNest} evidence. These evidence, compared to the case of no trends in RV, i.e. a constant RV, turned out to be statistically insignificant and we therefore can  exclude such trends with an amplitude larger than about 30\,m\,s$^{-1}$.

\subsection{Inclinations of the orbital planes} 

\begin{figure}[t]
	\begin{center}
	\includegraphics[width=0.5\textwidth]{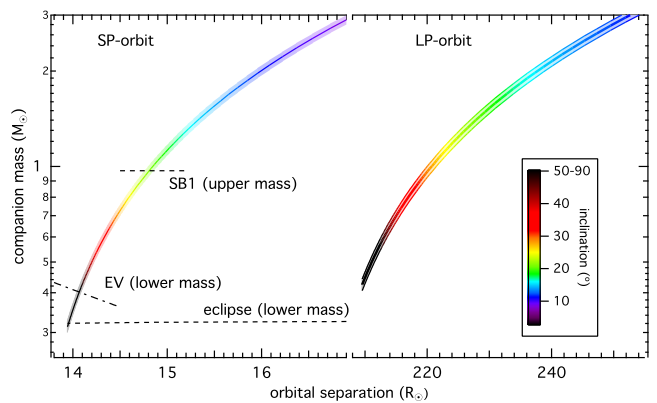}
	\caption{Range of companion mass' and orbital separations determined by orbital solutions as a function of the orbit inclination for the short-period companion (SP, left) and the long-period companion (LP, right), assuming a primary mass of 3.0$\pm$0.2\,M\sun . While the lower dashed line indicates the maximum inclination (and therefore minimum companion mass) to avoid an eclipse, the upper dashed line marks the upper mass limit coming from the fact that \hd\ is a SB1 system. The dash-dotted line gives the maximum inclinations the system can have to produce ellipsoidal light variations of no larger than 0.7\,ppt.}  
	\label{fig:massfunct} 
	\end{center} 
\end{figure}

Because we do not find eclipses or tidally induced flux modulations in the light curve\footnote{While the BRITE observations are not long enough to search for the photometric signature of the LP orbit, there is clearly no signal neither in the raw nor in the post-processed BTr data set}, which is larger than about 0.3\,ppt at the orbital period of the SP companion, it is not possible to directly infer the inclination ($i$) of the orbital planes from the available RV measurements. Hence we can only determine the projected semi-major axes and the mass functions (see Tab.\,\ref{tab:Borbit}): 
\begin{equation}
f_{(m)} = \frac{(M_{2,3} \sin i)^3}{(M_1 + M_{2,3})^2} = \frac{K_\mathrm{SP,LP}^3P_\mathrm{SP,LP}}{2\pi G} (1-e_\mathrm{SP,LP}^2)^{1.5},
\end{equation}
where the indices 1, 2, and 3 indicate the central star and SP and the LP companion, respectively. From seismology we have, however, a good idea about the mass of \hd\ and can therefore estimate the companion masses $M_2$ and $M_3$ as a function of $i$. From this we can now determine the orbital separations: 
\begin{equation}
a_\mathrm{SP,LP} = a_\mathrm{SP,LP,1} + a_{2,3} = \frac{K_\mathrm{SP,LP}P_\mathrm{SP,LP}}{2\pi}(1+q), 
\end{equation}
where $q$ is the mass ratio $M_1/M_{2,3}$. Considering all uncertainties and a conservative estimate of $M_1=3.0\pm0.2$M\sun\ the resulting companion masses and orbital separations are shown in Fig.\,\ref{fig:massfunct} as a function of the orbital inclination. In this context we can now test various assumptions. 

A possible limitation for the inclination comes from the fact that we do not see any sign of eclipses or rotational modulation in the BRITE photometry, e.g., due to a distorted geometry of a contact or semi-detached binary. For a given orbital separation ($a$), a limit for $i$ to avoid an eclipse is determined by  $\cos i < (R_1+R_2)/(a)$, where we assume $R_2 \propto M_2^\zeta$, with $\zeta$=0.57 and 0.8 for stars with a mass above and below 1\,M\sun , respectively \citep[e.g.][]{torres2010}. The resulting maximum inclination can be translated into a minimum $M_2$ for which the system does not eclipse. Fig.\,\ref{fig:massfunct} shows this limit as a function of the orbital separation in units of R\sun\  and it is obvious that the system can only eclipse for inclinations larger than about 85\degr .  

Another constraint comes from ellipsoidal variations in non-eclipsing close binary systems whose components are distorted by their mutual gravitation. According to \cite{Morris1985}, the peak-to-peak amplitude of the primarys' light variations $\mathcal{A}_1$ can be connected to the system parameters as $R_1^3\sin^2{i}/(qa^3) = 3.07\mathcal{A}_1(3-u_1)/[(\tau + 1)(15+u_1)]$, where $\mathcal{A}_1 = 2.5 \log{[(1+x)/(1-x)] }$ and $u_1$ and $\tau$ are the limb-darkening and gravity-darkening coefficients, respectively. For small amplitudes $x = \mathcal{A} (I_2/I_1 + 1)$, with $\mathcal{A}$ and $I_2/I_1$ being the semi-amplitude of the ellipsoidal variations and the intensity ratio of the two stars, respectively, in the BTr passband. For stars of similar temperature their luminosity ratio translates directly into an intensity ratio in a given passband, but for stars with significantly different temperatures the different spectral energy distributions need to be taken into account as well. For simplicity we only consider the effects of black-body radiation. For main-sequence stars $L \propto M^{3.9}$ and $T \propto M^{1-\zeta/2}$ (with $\zeta$ = 0.8 and 0.57 for $M$ larger and smaller than 1\,M\sun , respectively). The mass ratio then translates into an intensity ratio according to $I_2/I_1 = (M_2/M_1)^{3.9} E_{(\lambda,T_1)}/E_{(\lambda,T_2)}$, where $E_{(\lambda,T)}$ is the black-body radiation for a temperature $T$ at the central wavelength $\lambda$ of the passband. Instead of the original bolometric limb- and gravity-darkening coefficients used in \cite{Morris1985} we adopt $u_1 = 0.39$ and $\tau = 0.45$ from \cite{Claret2011}, which are computed for the SDSS $r'$ passband filter (which is similar to the BTr passband). Ellipsoidal variations usually have a double-peak structure in the light curve so that we can expect to find the strongest peak in the amplitude spectrum at twice the orbital frequency of 0.603644\d . We do indeed find a significant frequency at 0.60535\d\ ($f_{10}$ in Tab.\,\ref{tab:freq}), which differs by less than 2$\sigma$ from twice the orbital frequency. We, however, identify this as the $m=1$ component of the doublet $\nu_3$. Even though this interpretation seems more plausible we can most conservatively assume that the signal amplitude of about 0.7\,ppt is entirely due to ellipsoidal variations. Using the relation shown in Fig.\,\ref{fig:massfunct} we find a maximum inclination of 53\degr\ that the system can have to produce ellipsoidal light variations with an amplitude of less than 0.7\,ppt. This translates into a minimum companion mass and orbital separation of about 0.4\,M\sun\ and 14.1\,R\sun , respectively.

At first sight the spectroscopic observations of \hd\ are dominated by the spectral signature of the primary star with no obvious evidence for spectral lines originating from one of the companion stars. This indicates that the latter are much fainter than the primary star. For a more comprehensive testimony we search for the spectroscopic signature of the inner companion as follows. Its radial velocity is mirrored to the system velocity with an amplitude scaled by the mass ratio of the two stars. We therefore correct the observed spectra for mirrored radial velocities that result from a given mass ratio and average the spectra that were observed at phases around the RV maxima (i.e., $\pm$0.1 in phase, which covers ten spectra). If the companion were bright enough we would find spectral lines at the expected RV. Compared to looking at individual spectra this approach increases the sensitivity by averaging various spectra and also allows us to distinguish between weak lines of the central star and spectral lines originating from the companion. The absence of spectral lines either means that we did not get the mass ratio right or that the companion's spectral lines are too diluted to stand out of the noise. We tested companion masses from 0.3 to 3\,M\sun\ (in steps of 0.1) but did not find any signature of individual spectral lines from the companion in the wavelength range around the \ion{Fe}{II} line, as is illustrated in Fig.\,\ref{fig:lineprofiles}. This allows us to define an upper limit for the luminosity of the companion star. The averaged spectra (for each companion mass) scatter in normalised flux by about 0.0025 in the range where we expect a spectral line of the companion, which would then need to be at least 0.0075 deep to to be differentiated from the noise by more than 3$\sigma$. Ignoring white dwarfs for the time being, the companion is very likely a low-mass main sequence star very close to the ZAMS (given the relatively young age of \hd\ for a main sequence star). Given this we assume the intrinsic normalised depth of the genuine spectral line of the companion to be at least 0.5. This results in an intensity dilution factor of at least 67 at the given wavelength. As for the ellipsoidal variations, we also need to account for the stars' different spectral energy distributions to translate the intensity ratio into a mass ratio. Using mass and effective temperature of our representative model (see Tab.\,\ref{tab:model}) the companion mass needs to be below 0.97\,M\sun\ in order to be undetectable in our analysis. According to Fig.\,\ref{fig:massfunct} this translates into a minimum inclination of about 21\degr\ and a maximum orbital separation of about 14.8\,R\sun . The above considerations do not explicitly account for rotational broadening of the companions' spectral lines but already include some margin for moderate rotation (in fact the considered spectral lines in low-mass stars are much deeper than 0.5). However, even cutting the minimum dilution factor in half (i.e., allowing for more rotational broadening) does only increase the upper mass limit by about 20\%.

The above considerations are only valid for main sequence stars but the companion could also well be a white dwarf (WD). They are orders of magnitudes less luminous than \hd . It is therefore practically impossible to see the spectral signature of a white dwarf in a composite spectrum with a B-type star. In a statistical sense a WD-companion would have a mass of about 0.6\,M\sun\ \citep[e.g.][]{Tremblay2016} and very little chance to be more massive than one solar mass.

%--------------------------------------------------------------------
%-------------------------------------------------------------------- 

If we assume the outer layers of \hd\ to rotate with the orbital period of the SP companion (i.e., synchronised rotation), the equatorial rotational velocity would be of the order of $v_{rot} = 2\pi R_1/P_\mathrm{SP} \simeq 39.6$\,km/s. A comparison with the observed \vs\ = 6 -- 10\,km/s gives an inclination of 9 -- 15\degr. Based on the above (conservative) considerations we can, however, exclude such a low inclination. In fact, it would require a 1.5 -- 2.5\,M\sun\ companion, which we would definitely detect in the spectra. The obvious solutions for this contradiction are that either the rotation axis of \hd\ is not aligned with the SP orbital axis or its top-layer rotation is not synchronised. Even though very speculative, we remind that the doublet structure of the oscillation modes indicates that we observe the oscillations more equator-on than pole-on (i.e., the inclination of the pulsation axis is ``large''). Assuming the pulsation axis to be aligned with the rotation axis would then favour mis-alignment with the SP orbit. On the other hand, our analysis of the internal rotation profile (Sec.\,\ref{Sec:Rot}) indicates that the surface of \hd\ rotates with a period in the range of 2 -- 11\,d, and does therefore not exclude that the surface rotation is somewhat slower than the SP orbital period (i.e., the equatorial rotational velocity is smaller than for synchronised rotation and therefore the inclination is larger than 9 -- 15\degr). We think the latter explanation is more plausible. Given that the system is quite young, that circularisation time scales in binary systems are shorter than synchronisation time scales \citep[e.g.][]{zahn2013} and that the SP orbit is still not fully circularised ($e_\mathrm{SP} \ne 0$), it is plausible that the gravitational interaction between \hd\ and its SP companion is just about to finish circularisation. Hence,  spin-alignment is still active and the rotation of the main star is accelerating. A better understanding of this scenario would, however, require detailed modelling of the dynamical history of the multiple system, which is beyond the scope of this paper.

\section{Discussion}	\label{sec:discussion}

Rotation is a key process of the evolution of stars, that is still not fully understood. Along with magnetic fields, rotation strongly affects the transport of chemical elements and of angular momentum (AM), and therefore the structure and evolution of stars \citep[e.g.][]{Zahn1992,Maeder2012}. Isolated stars are believed to conserve their total AM throughout evolution from the main sequence to high up the giant branch. During this evolution, their cores (mostly) contract while their envelopes (mostly) expand until they are eventually ejected carrying away AM to the stellar environment. If stars would locally conserve AM throughout their life, their cores would spin up and the surviving compact objects would spin much faster than is actually observed \citep[e.g.][]{Kawaler2015}. This implies that standard formulations of stellar structure and evolution miss one (or more) essential processes. Which and what  their individual contributions are, as well as over which timescales they operate, is still unclear and none of the proposed mechanisms \citep[e.g.][]{Tayar2013,Cantiello2014,Fuller2014,rudiger2015} can sufficiently reproduce the observations. 

\begin{figure}[t]
	\begin{center}
	\includegraphics[width=0.5\textwidth]{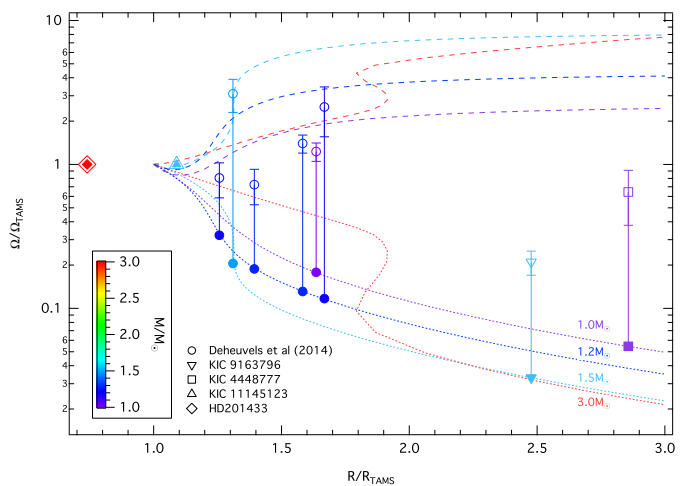}
	\caption{Mean core (dashed lines) and envelope (dotted lines) rotation rate during the evolution of YREC models (from the TAMS to the RGB) with various masses (colour coded) assuming local conservation of angular momentum and rigid rotation on the main sequence. The rotation rate and stellar radius are given relative their respective values on the TAMS. The filled symbols correspond to the relative envelope rotation rates of various stars with a given mass and radius. The core rotation rate (open symbols) is determined from this value and the oberved core-to-envelope rotation gradient.} 
	\label{fig:rotevo} 
	\end{center} 
\end{figure}

A first step to improve our understanding of stellar rotation is to provide a clear view on how AM is distributed inside stars as a function of various parameters (e.g., stellar mass) and during different evolutionary stages. Prime candidates for such studies are sub-giant and red-giant stars for which the contrast between the core and  surface rotation has been measured using rotationally split mixed p/g modes \citep[e.g.][]{Beck2012,Mosser2012}. These measurements are, however, only one piece of the puzzle as the star's rotational history remains unknown and it is thus difficult to compare the observations to model predictions. On the other hand, the outer envelope contains almost all of the star's AM. Based on the measured surface rotation rate of an evolved star one can thus well approximate how fast its surface was rotating when the star was still on the main sequence by simply following the internal mass distribution of a representative model backwards. By assuming rigid rotation on the MS the entire rotational history of an evolved star can then be normalised to its MS properties, which allows a direct comparison of different stars. We have done this for various stars in Fig.\,\ref{fig:rotevo}, where we show the core-to-surface rotation contrast as a function of the stellar radius for six sub-giants \citep{Deheuvels2014} and the more evolved red giants KIC\,4448777 \citep{DiMauro2016} and KIC\,9163796 (Beck et al., submitted). Once the diagram is more populated (and such analyses are underway) it will allow us to follow the AM redistribution in the interior of these stars and therefore provide the observational constraints necessary to test the various AM transport mechanisms. 

Our concept of ``visualising'' the AM transport in evolved stars depends critically on the assumption that stars rotate rigidly in first proximity when leaving the MS. This is in fact difficult to test observationally and has so far only been shown for three late A to early F-type stars (KIC\,11145123 -- \citealt{Kurtz2014}; KIC\,9244992 -- \citealt{Saio2015}; KIC\,7661054 -- \citealt{Murphy2016}). Even though \cite{triana2015} could not rule out rigid rotation for the B8V star KIC\,10526294, our analysis of \hd\ provides the first confirmation that this important assumption also holds true for the progenitors of more massive red giants.

The models used to place the observations in Fig.\,\ref{fig:rotevo} were computed with the Yale stellar evolution code \citep[YREC;][]{Guenther1992,Demarque2008} for near-solar composition and calibrated mixing length parameter ($Z = 0.02$, $Y = 0.27$, $\alpha_\mathrm{MLT} = 1.8$) assuming standard solar mixture \citep{Grevesse1996}.  More details about the constitutive physics are described by, e.g., \cite{Kallinger2012} and references therein.
  
%KIC 11145123 \citep{Kurtz2014}

%--------------------------------------------------------------------
%--------------------------------------------------------------------

\section{Summary}	\label{sec:summary}

\hd\ has the potential to be a Rosetta-stone SPB star. It is very bright and, hence, allows for detailed investigations with spectroscopic, polarimetric, interferometric and photometric observations on all relevant scales of temporal and photometric resolution, in the visible region and beyond. It is a success of BRITE-Constellation to have identified the potential of this object. 

\hd\ is a known single-lined spectroscopic triple system consisting of a massive SPB star orbited by two low-mass stars with periods of about 3.31 and 154\,d. Our 27 new high-resolution spectra, obtained in 2016, extended the time base to slightly more than 96 years with a total of 231 spectra usable for an orbital analysis. Such a long time base allowed us to check with Bayesian techniques for variability of the short-period companion, and, indeed, we find an annual relative period change rate of (3.2$\pm$1.0)$\times$10$^{-8}$, or that the orbital period is now about 0.9\,s shorter than about 10\,000 orbits earlier, in 1921. Even though we cannot exclude other explanations (like the influence of an undetected third companion) for this it appears plausible that the decreasing period is the signature of an angular momentum transfer in the inner binary system. The orbital elements are presented in Tab.\,\ref{tab:Borbit} and agree with \cite{barlow1989} within the errors, but are now more precise by a factor of 2 to 10. In addition, we also allow for elliptical orbits, but find the eccentricity of the inner companion to be small ($e_\mathrm{SP} = 0.015\pm0.003$) and close to the circular orbit, as assumed in pervious analyses. Because we do not find eclipses or tidally induced flux modulations in the BRITE observations and since the spectral signature of the companions is not visible in our spectroscopic time series we can limit the mass of the inner companion to the range of about 0.4 -- 1\,M\sun .

Obviously, pulsation effects in spectra interfere with the dynamical analysis of \hd\ and, hence, a photometric frequency analysis was performed contemporaneously. Indicated by peaks in the Fourier spectrum of the BRITE observations being broader than expected from the spectral window function, we applied a Bayesian-based technique to identify frequencies which are separated by less than the formal (Rayleigh-) frequency resolution of 1/$T$, with $T$ being in our case 156\,d. We can reliably distinguish between a single frequency and a pair of frequencies \citep{kal2016}, if the latter are separated by more than $\sim0.5/T$ and their amplitudes are sufficiently large, in our case $\sim$1\,ppt. Remarkably, the frequency errors are only slightly larger than for a mono-periodic signal, but rarely exceed $0.1/T$ (i.e., one tenth of the Rayleigh frequency resolution). 

The pulsation spectrum (see Fig.\,\ref{fig:fspec}) of \hd\ shows the strongest peaks between about 0.4\d\  and 1\d, and another group of peaks between about 1.5\d\ and 2\d. From our list of significant independent frequencies we can identify nine rotationally split doublets ($\nu_1$ -- $\nu_9$ in Fig.\,\ref{fig:freq}), which indicate that all modes have the same spherical degree of $l=1$ (Tab.\,\ref{tab:split}). The available spectra strongly indicate that the line profile variations are due to non-radial oscillations and that their morphology is consistent with $l=1$ modes. Assuming that some frequency pairs could not be identified in the BTr data due to insufficient SNR, the average period difference for the sequence of pairs is about 5030\,s, which is a typical value for a star like \hd . Our interpretation of the BTr photometry is fully consistent with the almost 8-year long but poorer-quality observations obtained by the \smei\ instrument on board the Coriolis satellite, for which we identify three rotationally split triplets whose central frequencies and splittings agree well with those extracted from the BTr data. We also find the amplitudes of the individual components to be modulated on timescales of about 1\,500\,d .

A highlight from the BRITE photometry of \hd\ is a trend of increasing frequency splittings towards higher periods, which implies a non-rigid internal rotation profile - an outstanding detection for the group of SPB stars and a key to tackle one of the big open questions in stellar evolutionary theory, the transport of angular momentum inside stars. For this investigation we computed about 5\,200 MESA models (see Fig.\ref{fig:HRD}) and their non-adiabatic pulsation modes with the GYRE code. Among those we search for a representative model that reproduces the observed frequencies best using classical $\chi^2$ techniques but also more reliable statistical methods. Our best fit model has a mass and radius of 3.05\,M\sun\ and 2.6\,R\sun\, respectively, and is about half way through its main-sequence lifetime. 

\hd\ is almost a twin star to KIC\,10526294 \citep{papics2014}. Apart from their age (\hd\ appears to be more evolved) they are very similar in mass, surface gravity, and effective temperature and even the measured rotational splittings are similar in value and trend. \cite{triana2015} investigated the internal rotation profile of KIC\,10526294 and found the rotation rate near the core-envelope boundary to be well constrained. Their seismic data are consistent with rigid rotation but a profile with counter-rotation within the envelope has a slight statistical advantage over constant rotation. Our analysis of the internal rotation profile is more conclusive and we can statistically rule out counter-rotation in \hd . 

The modes that are accessible to our analysis probe the radiative envelope of \hd\ from the boundary of the convective core (r\,$\simeq0.11R_{*}$) up to about $0.98R_{*}$ (Fig.\ref{fig:BV}). According to our analysis, the measured rotational splittings do not support differential rotation in the radiative envelope of \hd\ below a radius fraction of about 0.9, but are, on the other hand, inconsistent with the entire radiative envelope rotating at a constant rate. Our Bayesian analysis provides strong evidence for a slowly (292$\pm$76\,d) and rigidly rotating envelope, topped by a thin and significantly more rapidly rotating surface layer reaching down to about 96\% of the radius. It is interesting to mention that this rotation at the surface is compatible, within the given uncertainties, with the orbital period of the innermost companion of \hd , suggesting an angular momentum transfer from the companion to the outer envelope of \hd, causing its acceleration.

Combining the asteroseismic inferences with the spectroscopic measurement of the projected rotational velocity and our orbital analysis of the inner binary system we conclude that tidal interactions between the primary SPB star and its inner companion have almost circularised the orbit but not yet aligned all spins of the system and just started to accelerate the rotation of the SPB star to synchronise rotation. This makes the system quite interesting in order to study tidal interaction of binaries ``in action''.

We further refer to a detailed follow-up study about the atmosphere of \hd . The star's atypically sharp spectral lines make it a perfect candidate to study NLTE effects in the atmosphere of a hot star (Ryabchikova et al, in preparation).

\begin{acknowledgements}
We thank the technical team as well as the observers of the \Hermes spectrograph and Mercator Telescope, operated on the island of La Palma by the Flemish Community, at the Spanish Observatorio del Roque de los Muchachos of the Instituto de Astrofísica de Canarias. BTr operations are supported through a Canadian Space Agency (CSA) Academic Development grant. The Polish contribution to the BRITE mission is supported by the NCN grant 2011/01/M/ST9/05914. The research leading to these results has (partially) received funding from the European Research Council (ERC) under the European Union's Horizon 2020 research and innovation programme (grant agreement N$^{\circ}$670519: MAMSIE). TK and WW are grateful for funding via the Austrian Space Application Programme (ASAP) of the Austrian Research Promotion Agency (FFG) and BMVIT; and WW for support of the University of Vienna (IS 538001, IP 538007). PGB acknowledges the ANR (Agence Nationale de la Recherche, France) program IDEE (n$^\circ$ANR-12-BS05-0008) "Interaction Des Etoiles et des Exoplanetes”.  PGB also received funding from the CNES grants at CEA. YP and TR thank RFBR grant 15-02-06046 for partial financial support. TL acknowledges funding of the Austrian FFG within ASAP11 and by the grant WTZ CZ 15/2017. A. Pigulski and A. Popowicz acknowledge support from the Polish National Science Centre (grant No. 2016/21/B/ST9/01126 and 2016/21/D/ST9/00656). ES is thankful to the Russian Science Foundation (grant No. 14-50-00043) for financial support. GAW acknowledges Discovery Grant support from the Natural Sciences and Engineering Research Council (NSERC) of Canada. AFJM is grateful for financial aid from NSERC (Canada) and FQRNT (Quebec). GH acknowledges financial support by  the Polish NCN grant 2015/18/A/ST9/00578. KZ acknowledges support by the Austrian Fonds zur F\"orderung der wissenschaftlichen Forschung (FWF, project V431-NBL).

\end{acknowledgements}

\bibliographystyle{aa}
\bibliography{30625}

\begin{thebibliography}{96}
\expandafter\ifx\csname natexlab\endcsname\relax\def\natexlab#1{#1}\fi

\bibitem[{{Abt} \& {Cardona}(1984)}]{Abt1984}
{Abt}, H.~A. \& {Cardona}, O. 1984, \apj, 276, 266

\bibitem[{{Aerts}(2015)}]{aerts2015}
{Aerts}, C. 2015, in IAU Symposium, Vol. 307, New Windows on Massive Stars, ed.
  G.~{Meynet}, C.~{Georgy}, J.~{Groh}, \& P.~{Stee}, 154--164

\bibitem[{{Aerts} {et~al.}(2010){Aerts}, {Christensen-Dalsgaard}, \&
  {Kurtz}}]{aerts2010}
{Aerts}, C., {Christensen-Dalsgaard}, J., \& {Kurtz}, D.~W. 2010,
  {Asteroseismology}

\bibitem[{{Aerts} {et~al.}(2006){Aerts}, {De Cat}, {Kuschnig}, {Matthews},
  {Guenther}, {Moffat}, {Rucinski}, {Sasselov}, {Walker}, \&
  {Weiss}}]{aerts2006}
{Aerts}, C., {De Cat}, P., {Kuschnig}, R., {et~al.} 2006, \apjl, 642, L165

\bibitem[{{Aerts} {et~al.}(2003){Aerts}, {Thoul}, {Daszy{\'n}ska}, {Scuflaire},
  {Waelkens}, {Dupret}, {Niemczura}, \& {Noels}}]{aerts2003}
{Aerts}, C., {Thoul}, A., {Daszy{\'n}ska}, J., {et~al.} 2003, Science, 300,
  1926

\bibitem[{{Barlow}(1989)}]{barlow1989}
{Barlow}, D.~J. 1989, The Observatory, 109, 225

\bibitem[{{Beck} {et~al.}(2014){Beck}, {Hambleton}, {Vos}, {Kallinger},
  {Bloemen}, {Tkachenko}, {Garc{\'{\i}}a}, {{\O}stensen}, {Aerts}, {Kurtz}, {De
  Ridder}, {Hekker}, {Pavlovski}, {Mathur}, {De Smedt}, {Derekas}, {Corsaro},
  {Mosser}, {Van Winckel}, {Huber}, {Degroote}, {Davies}, {Pr{\v s}a},
  {Debosscher}, {Elsworth}, {Nemeth}, {Siess}, {Schmid}, {P{\'a}pics}, {de
  Vries}, {van Marle}, {Marcos-Arenal}, \& {Lobel}}]{Beck2014}
{Beck}, P.~G., {Hambleton}, K., {Vos}, J., {et~al.} 2014, \aap, 564, A36

\bibitem[{{Beck} {et~al.}(2012){Beck}, {Montalban}, {Kallinger}, {De Ridder},
  {Aerts}, {Garc{\'{\i}}a}, {Hekker}, {Dupret}, {Mosser}, {Eggenberger},
  {Stello}, {Elsworth}, {Frandsen}, {Carrier}, {Hillen}, {Gruberbauer},
  {Christensen-Dalsgaard}, {Miglio}, {Valentini}, {Bedding}, {Kjeldsen},
  {Girouard}, {Hall}, \& {Ibrahim}}]{Beck2012}
{Beck}, P.~G., {Montalban}, J., {Kallinger}, T., {et~al.} 2012, \nat, 481, 55

\bibitem[{{Breger} {et~al.}(1993){Breger}, {Stich}, {Garrido}, {Martin},
  {Jiang}, {Li}, {Hube}, {Ostermann}, {Paparo}, \& {Scheck}}]{breger1993}
{Breger}, M., {Stich}, J., {Garrido}, R., {et~al.} 1993, \aap, 271, 482

\bibitem[{{Cantiello} {et~al.}(2014){Cantiello}, {Mankovich}, {Bildsten},
  {Christensen-Dalsgaard}, \& {Paxton}}]{Cantiello2014}
{Cantiello}, M., {Mankovich}, C., {Bildsten}, L., {Christensen-Dalsgaard}, J.,
  \& {Paxton}, B. 2014, \apj, 788, 93

\bibitem[{{Christensen-Dalsgaard}(1990)}]{Dalsgaard1990}
{Christensen-Dalsgaard}, J. 1990, in Reviews in Modern Astronomy, Vol.~3,
  Reviews in Modern Astronomy, ed. G.~{Klare}, 313--349

\bibitem[{{Claret} \& {Bloemen}(2011)}]{Claret2011}
{Claret}, A. \& {Bloemen}, S. 2011, \aap, 529, A75

\bibitem[{{Cox}(1980)}]{Cox1980}
{Cox}, J.~P. 1980, {Theory of stellar pulsation}

\bibitem[{{Daszy{\'n}ska-Daszkiewicz}(2008)}]{2008CoAst.152..140D}
{Daszy{\'n}ska-Daszkiewicz}, J. 2008, Communications in Asteroseismology, 152,
  140

\bibitem[{{De Cat} \& {Aerts}(2002)}]{DeCat2002}
{De Cat}, P. \& {Aerts}, C. 2002, \aap, 393, 965

\bibitem[{{Deheuvels} {et~al.}(2014){Deheuvels}, {Do{\u g}an}, {Goupil},
  {Appourchaux}, {Benomar}, {Bruntt}, {Campante}, {Casagrande}, {Ceillier},
  {Davies}, {De Cat}, {Fu}, {Garc{\'{\i}}a}, {Lobel}, {Mosser}, {Reese},
  {Regulo}, {Schou}, {Stahn}, {Thygesen}, {Yang}, {Chaplin},
  {Christensen-Dalsgaard}, {Eggenberger}, {Gizon}, {Mathis},
  {Molenda-{\.Z}akowicz}, \& {Pinsonneault}}]{Deheuvels2014}
{Deheuvels}, S., {Do{\u g}an}, G., {Goupil}, M.~J., {et~al.} 2014, \aap, 564,
  A27

\bibitem[{{Demarque} {et~al.}(2008){Demarque}, {Guenther}, {Li}, {Mazumdar}, \&
  {Straka}}]{Demarque2008}
{Demarque}, P., {Guenther}, D.~B., {Li}, L.~H., {Mazumdar}, A., \& {Straka},
  C.~W. 2008, \apss, 316, 31

\bibitem[{{Di Mauro} {et~al.}(2016){Di Mauro}, {Ventura}, {Cardini}, {Stello},
  {Christensen-Dalsgaard}, {Dziembowski}, {Patern{\`o}}, {Beck}, {Bloemen},
  {Davies}, {De Smedt}, {Elsworth}, {Garc{\'{\i}}a}, {Hekker}, {Mosser}, \&
  {Tkachenko}}]{DiMauro2016}
{Di Mauro}, M.~P., {Ventura}, R., {Cardini}, D., {et~al.} 2016, \apj, 817, 65

\bibitem[{{Dupret} {et~al.}(2004){Dupret}, {Thoul}, {Scuflaire},
  {Daszy{\'n}ska-Daszkiewicz}, {Aerts}, {Bourge}, {Waelkens}, \&
  {Noels}}]{dupret2004}
{Dupret}, M.-A., {Thoul}, A., {Scuflaire}, R., {et~al.} 2004, \aap, 415, 251

\bibitem[{{Dziembowski} {et~al.}(1993){Dziembowski}, {Moskalik}, \&
  {Pamyatnykh}}]{Dziembowski1993}
{Dziembowski}, W.~A., {Moskalik}, P., \& {Pamyatnykh}, A.~A. 1993, \mnras, 265,
  588

\bibitem[{Eggleton(2006)}]{Eggleton2006}
Eggleton, P. 2006, Cambridge Astrophysics Series, Vol.~40, Evolutionary
  Processes in Binary and Multiple Stars (Cambridge; New York: Cambridge
  University Press)

\bibitem[{{Eyles} {et~al.}(2003){Eyles}, {Simnett}, {Cooke}, {Jackson},
  {Buffington}, {Hick}, {Waltham}, {King}, {Anderson}, \&
  {Holladay}}]{Eyles2003}
{Eyles}, C.~J., {Simnett}, G.~M., {Cooke}, M.~P., {et~al.} 2003, \solphys, 217,
  319

\bibitem[{{Feroz} {et~al.}(2009){Feroz}, {Hobson}, \& {Bridges}}]{feroz2009}
{Feroz}, F., {Hobson}, M.~P., \& {Bridges}, M. 2009, \mnras, 398, 1601

\bibitem[{{Fuller} {et~al.}(2014){Fuller}, {Lecoanet}, {Cantiello}, \&
  {Brown}}]{Fuller2014}
{Fuller}, J., {Lecoanet}, D., {Cantiello}, M., \& {Brown}, B. 2014, \apj, 796,
  17

\bibitem[{{Gieseking} \& {Seggewiss}(1978)}]{gieseking1978}
{Gieseking}, F. \& {Seggewiss}, W. 1978, \aap, 68, 437

\bibitem[{{Gizon} \& {Solanki}(2003)}]{gizon2003}
{Gizon}, L. \& {Solanki}, S.~K. 2003, \apj, 589, 1009

\bibitem[{{Grevesse} {et~al.}(1996){Grevesse}, {Noels}, \&
  {Sauval}}]{Grevesse1996}
{Grevesse}, N., {Noels}, A., \& {Sauval}, A.~J. 1996, in Astronomical Society
  of the Pacific Conference Series, Vol.~99, Cosmic Abundances, ed. S.~S.
  {Holt} \& G.~{Sonneborn}, 117

\bibitem[{{Gruber} {et~al.}(2012){Gruber}, {Saio}, {Kuschnig}, {Fossati},
  {Handler}, {Zwintz}, {Weiss}, {Matthews}, {Guenther}, {Moffat}, {Rucinski},
  \& {Sasselov}}]{gruber2012}
{Gruber}, D., {Saio}, H., {Kuschnig}, R., {et~al.} 2012, \mnras, 420, 291

\bibitem[{{Gruberbauer} {et~al.}(2009){Gruberbauer}, {Kallinger}, {Weiss}, \&
  {Guenther}}]{Gruberbauer2009}
{Gruberbauer}, M., {Kallinger}, T., {Weiss}, W.~W., \& {Guenther}, D.~B. 2009,
  \aap, 506, 1043

\bibitem[{{Guenther} \& {Brown}(2004)}]{guenther2004}
{Guenther}, D.~B. \& {Brown}, K.~I.~T. 2004, \apj, 600, 419

\bibitem[{{Guenther} {et~al.}(1992){Guenther}, {Demarque}, {Kim}, \&
  {Pinsonneault}}]{Guenther1992}
{Guenther}, D.~B., {Demarque}, P., {Kim}, Y.-C., \& {Pinsonneault}, M.~H. 1992,
  \apj, 387, 372

\bibitem[{{Guthnik}(1938)}]{guthnik1938}
{Guthnik}, P. 1938, Abh. Preuss. Akad. Wissenschaften, 3

\bibitem[{{Guthnik}(1939)}]{guthnik1939}
{Guthnik}, P. 1939, Abh. Preuss. Akad. Wissenschaften, 6

\bibitem[{{Guthnik}(1942)}]{guthnik1942}
{Guthnik}, P. 1942, Abh. Preuss. Akad. Wissenschaften, 7

\bibitem[{{Hick} {et~al.}(2007){Hick}, {Buffington}, \& {Jackson}}]{Hick2007}
{Hick}, P., {Buffington}, A., \& {Jackson}, B.~V. 2007, Proc. SPIE, 6689,
  66890C

\bibitem[{{Hoffleit}(1977)}]{Hoffleit1977}
{Hoffleit}, D. 1977, Information Bulletin on Variable Stars, 1283

\bibitem[{{Jackson} {et~al.}(2004){Jackson}, {Buffington}, {Hick}, {Altrock},
  {Figueroa}, {Holladay}, {Johnston}, {Kahler}, {Mozer}, {Price}, {Radick},
  {Sagalyn}, {Sinclair}, {Simnett}, {Eyles}, {Cooke}, {Tappin}, {Kuchar},
  {Mizuno}, {Webb}, {Anderson}, {Keil}, {Gold}, \& {Waltham}}]{Jackson2004}
{Jackson}, B.~V., {Buffington}, A., {Hick}, P.~P., {et~al.} 2004, \solphys,
  225, 177

\bibitem[{Jeffreys(1998)}]{jeffreys98}
Jeffreys, H. 1998, Theory of probability, Oxford Classic Texts in the Physical
  Sciences (New York: The Clarendon Press Oxford University Press), xii+459,
  reprint of the 1983 edition

\bibitem[{{Jerzykiewicz} {et~al.}(2013){Jerzykiewicz}, {Lehmann}, {Niemczura},
  {Molenda-{\.Z}akowicz}, {Dymitrov}, {Fagas}, {Guenther}, {Hartmann},
  {Hrudkov{\'a}}, {Kami{\'n}ski}, {Moffat}, {Kuschnig}, {Leto}, {Matthews},
  {Rowe}, {Ruci{\'n}ski}, {Sasselov}, \& {Weiss}}]{Jerzykiewicz2013}
{Jerzykiewicz}, M., {Lehmann}, H., {Niemczura}, E., {et~al.} 2013, \mnras, 432,
  1032

\bibitem[{{Kallinger} {et~al.}(2010){Kallinger}, {Gruberbauer}, {Guenther},
  {Fossati}, \& {Weiss}}]{kallinger2010}
{Kallinger}, T., {Gruberbauer}, M., {Guenther}, D.~B., {Fossati}, L., \&
  {Weiss}, W.~W. 2010, \aap, 510, A106

\bibitem[{{Kallinger} {et~al.}(2012){Kallinger}, {Hekker}, {Mosser}, {De
  Ridder}, {Bedding}, {Elsworth}, {Gruberbauer}, {Guenther}, {Stello}, {Basu},
  {Garc{\'{\i}}a}, {Chaplin}, {Mullally}, {Still}, \&
  {Thompson}}]{Kallinger2012}
{Kallinger}, T., {Hekker}, S., {Mosser}, B., {et~al.} 2012, \aap, 541, A51

\bibitem[{{Kallinger} {et~al.}(2008){Kallinger}, {Reegen}, \&
  {Weiss}}]{kal2008}
{Kallinger}, T., {Reegen}, P., \& {Weiss}, W.~W. 2008, \aap, 481, 571

\bibitem[{{Kallinger} \& {Weiss}(2016)}]{kal2016}
{Kallinger}, T. \& {Weiss}, W.~W. 2016, ArXiv:1611.08167

\bibitem[{{Kawaler}(2015)}]{Kawaler2015}
{Kawaler}, S.~D. 2015, in Astronomical Society of the Pacific Conference
  Series, Vol. 493, 19th European Workshop on White Dwarfs, ed. P.~{Dufour},
  P.~{Bergeron}, \& G.~{Fontaine}, 65

\bibitem[{{Kopp} \& {Lean}(2011)}]{kopp2011}
{Kopp}, G. \& {Lean}, J.~L. 2011, \grl, 38, L01706

\bibitem[{{Kurtz} {et~al.}(2014){Kurtz}, {Saio}, {Takata}, {Shibahashi},
  {Murphy}, \& {Sekii}}]{Kurtz2014}
{Kurtz}, D.~W., {Saio}, H., {Takata}, M., {et~al.} 2014, \mnras, 444, 102

\bibitem[{{Kuschnig} {et~al.}(1997){Kuschnig}, {Weiss}, {Gruber}, {Bely}, \&
  {Jenkner}}]{kuschnig1997}
{Kuschnig}, R., {Weiss}, W.~W., {Gruber}, R., {Bely}, P.~Y., \& {Jenkner}, H.
  1997, \aap, 328, 544

\bibitem[{{Lejeune} \& {Schaerer}(2001)}]{Lejeune2001}
{Lejeune}, T. \& {Schaerer}, D. 2001, \aap, 366, 538

\bibitem[{{Lohr} {et~al.}(2012){Lohr}, {Norton}, {Kolb}, {Anderson}, {Faedi},
  \& {West}}]{lohr2012}
{Lohr}, M.~E., {Norton}, A.~J., {Kolb}, U.~C., {et~al.} 2012, \aap, 542, A124

\bibitem[{{Loumos} \& {Deeming}(1978)}]{Loumos1978}
{Loumos}, G.~L. \& {Deeming}, T.~J. 1978, \apss, 56, 285

\bibitem[{{Maeder} \& {Meynet}(2012)}]{Maeder2012}
{Maeder}, A. \& {Meynet}, G. 2012, Reviews of Modern Physics, 84, 25

\bibitem[{{Matthews} {et~al.}(2004){Matthews}, {Kuschnig}, {Guenther},
  {Walker}, {Moffat}, {Rucinski}, {Sasselov}, \& {Weiss}}]{matthews2004}
{Matthews}, J.~M., {Kuschnig}, R., {Guenther}, D.~B., {et~al.} 2004, \nat, 430,
  51

\bibitem[{{Meisel}(1968)}]{Meisel1968}
{Meisel}, D.~D. 1968, \aj, 73, 350

\bibitem[{{Miglio} {et~al.}(2008){Miglio}, {Montalb{\'a}n}, {Noels}, \&
  {Eggenberger}}]{miglio2008}
{Miglio}, A., {Montalb{\'a}n}, J., {Noels}, A., \& {Eggenberger}, P. 2008,
  \mnras, 386, 1487

\bibitem[{{Mittermayer} \& {Weiss}(2003)}]{Mittermayer2003}
{Mittermayer}, P. \& {Weiss}, W.~W. 2003, \aap, 407, 1097

\bibitem[{{Moravveji} {et~al.}(2015){Moravveji}, {Aerts}, {P{\'a}pics},
  {Triana}, \& {Vandoren}}]{Moravveji2015}
{Moravveji}, E., {Aerts}, C., {P{\'a}pics}, P.~I., {Triana}, S.~A., \&
  {Vandoren}, B. 2015, \aap, 580, A27

\bibitem[{{Morris}(1985)}]{Morris1985}
{Morris}, S.~L. 1985, \apj, 295, 143

\bibitem[{{Mosser} {et~al.}(2012){Mosser}, {Goupil}, {Belkacem}, {Marques},
  {Beck}, {Bloemen}, {De Ridder}, {Barban}, {Deheuvels}, {Elsworth}, {Hekker},
  {Kallinger}, {Ouazzani}, {Pinsonneault}, {Samadi}, {Stello}, {Garc{\'{\i}}a},
  {Klaus}, {Li}, {Mathur}, \& {Morris}}]{Mosser2012}
{Mosser}, B., {Goupil}, M.~J., {Belkacem}, K., {et~al.} 2012, \aap, 548, A10

\bibitem[{{Murphy} {et~al.}(2016){Murphy}, {Fossati}, {Bedding}, {Saio},
  {Kurtz}, {Grassitelli}, \& {Wang}}]{Murphy2016}
{Murphy}, S.~J., {Fossati}, L., {Bedding}, T.~R., {et~al.} 2016, \mnras, 459,
  1201

\bibitem[{{Pablo} {et~al.}(2016){Pablo}, {Whittaker}, {Popowicz}, {Mochnacki},
  {Kuschnig}, {Grant}, {Moffat}, {Rucinski}, {Matthews},
  {Schwarzenberg-Czerny}, {Handler}, {Weiss}, {Baade}, {Wade},
  {Zoc{\l}o{\'n}ska}, {Ramiaramanantsoa}, {Unterberger}, {Zwintz}, {Pigulski},
  {Rowe}, {Koudelka}, {Orlea{\'n}ski}, {Pamyatnykh}, {Neiner}, {Wawrzaszek},
  {Marciniszyn}, {Romano}, {Wo{\'z}niak}, {Zawistowski}, \& {Zee}}]{pablo2016}
{Pablo}, H., {Whittaker}, G.~N., {Popowicz}, A., {et~al.} 2016, \pasp, 128,
  125001

\bibitem[{{Pamyatnykh} {et~al.}(1998){Pamyatnykh}, {Dziembowski}, {Handler}, \&
  {Pikall}}]{Pamyatnykh1998}
{Pamyatnykh}, A.~A., {Dziembowski}, W.~A., {Handler}, G., \& {Pikall}, H. 1998,
  \aap, 333, 141

\bibitem[{{P{\'a}pics} {et~al.}(2014){P{\'a}pics}, {Moravveji}, {Aerts},
  {Tkachenko}, {Triana}, {Bloemen}, \& {Southworth}}]{papics2014}
{P{\'a}pics}, P.~I., {Moravveji}, E., {Aerts}, C., {et~al.} 2014, \aap, 570, C4

\bibitem[{{P{\'a}pics} {et~al.}(2015){P{\'a}pics}, {Tkachenko}, {Aerts}, {Van
  Reeth}, {De Smedt}, {Hillen}, {{\O}stensen}, \& {Moravveji}}]{papics2015}
{P{\'a}pics}, P.~I., {Tkachenko}, A., {Aerts}, C., {et~al.} 2015, \apjl, 803,
  L25

\bibitem[{{Paxton} {et~al.}(2011){Paxton}, {Bildsten}, {Dotter}, {Herwig},
  {Lesaffre}, \& {Timmes}}]{paxton2011}
{Paxton}, B., {Bildsten}, L., {Dotter}, A., {et~al.} 2011, \apjs, 192, 3

\bibitem[{{Paxton} {et~al.}(2013){Paxton}, {Cantiello}, {Arras}, {Bildsten},
  {Brown}, {Dotter}, {Mankovich}, {Montgomery}, {Stello}, {Timmes}, \&
  {Townsend}}]{paxton2013}
{Paxton}, B., {Cantiello}, M., {Arras}, P., {et~al.} 2013, \apjs, 208, 4

\bibitem[{{Pigulski} {et~al.}(2016){Pigulski}, {Cugier}, {Popowicz},
  {Kuschnig}, {Moffat}, {Rucinski}, {Schwarzenberg-Czerny}, {Weiss}, {Handler},
  {Wade}, {Koudelka}, {Matthews}, {Mochnacki}, {Orlea{\'n}ski}, {Pablo},
  {Ramiaramanantsoa}, {Whittaker}, {Zoc{\l}o{\'n}ska}, \&
  {Zwintz}}]{pigulski2016}
{Pigulski}, A., {Cugier}, H., {Popowicz}, A., {et~al.} 2016, \aap, 588, A55

\bibitem[{{Piskunov} \& {Valenti}(2016)}]{2016arXiv160606073P}
{Piskunov}, N. \& {Valenti}, J.~A. 2016, \aap, 597, A16

\bibitem[{{Popowicz} {et~al.}(2017){Popowicz}, {Bernacki}, \&
  {Pigulski}}]{popowicz2016}
{Popowicz}, A., {Bernacki}, K., \& {Pigulski}, A. 2017, A\&A (submitted)

\bibitem[{{Raskin}(2011)}]{RaskinPhD}
{Raskin}, G. 2011, PhD thesis, Institute of Astronomy, Katholieke Universiteit
  Leuven, Belgium

\bibitem[{{Raskin} {et~al.}(2011){Raskin}, {van Winckel}, {Hensberge},
  {Jorissen}, {Lehmann}, {Waelkens}, {Avila}, {de Cuyper}, {Degroote},
  {Dubosson}, {Dumortier}, {Fr{\'e}mat}, {Laux}, {Michaud}, {Morren}, {Perez
  Padilla}, {Pessemier}, {Prins}, {Smolders}, {van Eck}, \&
  {Winkler}}]{Raskin2011}
{Raskin}, G., {van Winckel}, H., {Hensberge}, H., {et~al.} 2011, A\&A, 526, A69

\bibitem[{{R{\"u}diger} {et~al.}(2015){R{\"u}diger}, {Gellert}, {Spada}, \&
  {Tereshin}}]{rudiger2015}
{R{\"u}diger}, G., {Gellert}, M., {Spada}, F., \& {Tereshin}, I. 2015, \aap,
  573, A80

\bibitem[{{Ryabchikova} {et~al.}(2015{\natexlab{a}}){Ryabchikova}, {Piskunov},
  {Kurucz}, {Stempels}, {Heiter}, {Pakhomov}, \&
  {Barklem}}]{2015PhyS...90e4005R}
{Ryabchikova}, T., {Piskunov}, N., {Kurucz}, R.~L., {et~al.}
  2015{\natexlab{a}}, \physscr, 90, 054005

\bibitem[{{Ryabchikova} {et~al.}(2016){Ryabchikova}, {Piskunov}, {Pakhomov},
  {Tsymbal}, {Titarenko}, {Sitnova}, {Alexeeva}, {Fossati}, \&
  {Mashonkina}}]{2016MNRAS.456.1221R}
{Ryabchikova}, T., {Piskunov}, N., {Pakhomov}, Y., {et~al.} 2016, \mnras, 456,
  1221

\bibitem[{{Ryabchikova} {et~al.}(2015{\natexlab{b}}){Ryabchikova}, {Piskunov},
  \& {Shulyak}}]{2015ASPC..494..308R}
{Ryabchikova}, T., {Piskunov}, N., \& {Shulyak}, D. 2015{\natexlab{b}}, in
  Astronomical Society of the Pacific Conference Series, Vol. 494, Physics and
  Evolution of Magnetic and Related Stars, ed. Y.~Y. {Balega}, I.~I.
  {Romanyuk}, \& D.~O. {Kudryavtsev}, 308

\bibitem[{{Ryabchikova} {et~al.}(2007){Ryabchikova}, {Sachkov}, {Kochukhov}, \&
  {Lyashko}}]{2007A&A...473..907R}
{Ryabchikova}, T., {Sachkov}, M., {Kochukhov}, O., \& {Lyashko}, D. 2007, \aap,
  473, 907

\bibitem[{{Sachkov} {et~al.}(2008){Sachkov}, {Kochukhov}, {Ryabchikova},
  {Huber}, {Leone}, {Bagnulo}, \& {Weiss}}]{2008MNRAS.389..903S}
{Sachkov}, M., {Kochukhov}, O., {Ryabchikova}, T., {et~al.} 2008, \mnras, 389,
  903

\bibitem[{{Saio} {et~al.}(2015){Saio}, {Kurtz}, {Takata}, {Shibahashi},
  {Murphy}, {Sekii}, \& {Bedding}}]{Saio2015}
{Saio}, H., {Kurtz}, D.~W., {Takata}, M., {et~al.} 2015, \mnras, 447, 3264

\bibitem[{{Schou} {et~al.}(1998){Schou}, {Antia}, {Basu}, {Bogart}, {Bush},
  {Chitre}, {Christensen-Dalsgaard}, {Di Mauro}, {Dziembowski}, {Eff-Darwich},
  {Gough}, {Haber}, {Hoeksema}, {Howe}, {Korzennik}, {Kosovichev}, {Larsen},
  {Pijpers}, {Scherrer}, {Sekii}, {Tarbell}, {Title}, {Thompson}, \&
  {Toomre}}]{Schou1998}
{Schou}, J., {Antia}, H.~M., {Basu}, S., {et~al.} 1998, \apj, 505, 390

\bibitem[{{Shulyak} {et~al.}(2004){Shulyak}, {Tsymbal}, {Ryabchikova},
  {St{\"u}tz}, \& {Weiss}}]{2004AA...428..993S}
{Shulyak}, D., {Tsymbal}, V., {Ryabchikova}, T., {St{\"u}tz}, C., \& {Weiss},
  W.~W. 2004, \aap, 428, 993

\bibitem[{{Sybilski} {et~al.}(2013){Sybilski}, {Konacki}, {Koz{\l}owski}, \&
  {He{\l}miniak}}]{Sybilski2013}
{Sybilski}, P., {Konacki}, M., {Koz{\l}owski}, S.~K., \& {He{\l}miniak}, K.~G.
  2013, \mnras, 431, 2024

\bibitem[{{Takeda} {et~al.}(2014){Takeda}, {Kawanomoto}, \&
  {Ohishi}}]{takeda2014}
{Takeda}, Y., {Kawanomoto}, S., \& {Ohishi}, N. 2014, \pasj, 66, 23

\bibitem[{{Tayar} \& {Pinsonneault}(2013)}]{Tayar2013}
{Tayar}, J. \& {Pinsonneault}, M.~H. 2013, \apjl, 775, L1

\bibitem[{{Tkachenko} {et~al.}(2012){Tkachenko}, {Lehmann}, {Smalley},
  {Debosscher}, \& {Aerts}}]{2012MNRAS.422.2960T}
{Tkachenko}, A., {Lehmann}, H., {Smalley}, B., {Debosscher}, J., \& {Aerts}, C.
  2012, \mnras, 422, 2960

\bibitem[{{Torres} {et~al.}(2010){Torres}, {Andersen}, \&
  {Gim{\'e}nez}}]{torres2010}
{Torres}, G., {Andersen}, J., \& {Gim{\'e}nez}, A. 2010, \aapr, 18, 67

\bibitem[{{Townsend} \& {Teitler}(2013)}]{townsend2013}
{Townsend}, R.~H.~D. \& {Teitler}, S.~A. 2013, \mnras, 435, 3406

\bibitem[{{Tremblay} {et~al.}(2016){Tremblay}, {Cummings}, {Kalirai},
  {G{\"a}nsicke}, {Gentile-Fusillo}, \& {Raddi}}]{Tremblay2016}
{Tremblay}, P.-E., {Cummings}, J., {Kalirai}, J.~S., {et~al.} 2016, \mnras,
  461, 2100

\bibitem[{{Triana} {et~al.}(2015){Triana}, {Moravveji}, {P{\'a}pics}, {Aerts},
  {Kawaler}, \& {Christensen-Dalsgaard}}]{triana2015}
{Triana}, S.~A., {Moravveji}, E., {P{\'a}pics}, P.~I., {et~al.} 2015, \apj,
  810, 16

\bibitem[{{Valenti} \& {Piskunov}(1996)}]{1996AAS..118..595V}
{Valenti}, J.~A. \& {Piskunov}, N. 1996, \aaps, 118, 595

\bibitem[{{van Leeuwen}(2007)}]{leeuwen2007}
{van Leeuwen}, F. 2007, \aap, 474, 653

\bibitem[{{Waelkens}(1991)}]{Waelkens1991}
{Waelkens}, C. 1991, \aap, 246, 453

\bibitem[{{Walker} {et~al.}(2003){Walker}, {Matthews}, {Kuschnig}, {Johnson},
  {Rucinski}, {Pazder}, {Burley}, {Walker}, {Skaret}, {Zee}, {Grocott},
  {Carroll}, {Sinclair}, {Sturgeon}, \& {Harron}}]{walker2003}
{Walker}, G., {Matthews}, J., {Kuschnig}, R., {et~al.} 2003, \pasp, 115, 1023

\bibitem[{{Walker} {et~al.}(2005){Walker}, {Kuschnig}, {Matthews}, {Cameron},
  {Saio}, {Lee}, {Kambe}, {Masuda}, {Guenther}, {Moffat}, {Rucinski},
  {Sasselov}, \& {Weiss}}]{walker2005}
{Walker}, G.~A.~H., {Kuschnig}, R., {Matthews}, J.~M., {et~al.} 2005, \apjl,
  635, L77

\bibitem[{{Weiss} {et~al.}(2014){Weiss}, {Rucinski}, {Moffat},
  {Schwarzenberg-Czerny}, {Koudelka}, {Grant}, {Zee}, {Kuschnig}, {Mochnacki},
  {Matthews}, {Orleanski}, {Pamyatnykh}, {Pigulski}, {Alves}, {Guedel},
  {Handler}, {Wade}, \& {Zwintz}}]{weiss2014}
{Weiss}, W.~W., {Rucinski}, S.~M., {Moffat}, A.~F.~J., {et~al.} 2014, \pasp,
  126, 573

\bibitem[{{Young}(1921)}]{young1921}
{Young}, R.~K. 1921, Publications of the Dominion Astrophysical Observatory
  Victoria, 1, 319

\bibitem[{{Zahn}(1992)}]{Zahn1992}
{Zahn}, J.-P. 1992, \aap, 265, 115

\bibitem[{{Zahn}(2013)}]{zahn2013}
{Zahn}, J.-P. 2013, in Lecture Notes in Physics, Berlin Springer Verlag, Vol.
  861, Lecture Notes in Physics, Berlin Springer Verlag, ed. J.~{Souchay},
  S.~{Mathis}, \& T.~{Tokieda}, 301

\end{thebibliography}

\end{document}